\documentclass[12pt,journal,onecolumn,english,draftcls]{IEEEtran}
\usepackage[T1]{fontenc}
\usepackage{amsthm}
\usepackage{amssymb}
\usepackage{graphicx}
\usepackage{bbm}
\makeatletter
\theoremstyle{plain}
\newtheorem{thm}{\protect\theoremname}
\theoremstyle{plain}
\newtheorem{cor}[thm]{\protect\corollaryname}
\theoremstyle{definition}
\newtheorem{defn}[thm]{\protect\definitionname}
\theoremstyle{plain}
\newtheorem{lem}[thm]{\protect\lemmaname}

\usepackage{subfigure}

\usepackage{footnote}
\makesavenoteenv{tabular}

\usepackage{lettrine}
\usepackage{hyperref}

\usepackage{amsthm}

\usepackage{amsmath}
\def\figref#1{Fig.\,\ref{#1}}

\usepackage{algorithmic}
\usepackage{float}
\newfloat{algorithm}{t}{lop}

\usepackage{cite}
\usepackage{url}

\newtheorem{definition}{Definition}

\newlength{\figwidth}
\ifCLASSOPTIONonecolumn
\else
\fi

\usepackage{breqn}
\usepackage{graphicx}
\usepackage{epstopdf}
\usepackage{amssymb}
\usepackage{amsmath}
\usepackage[misc,geometry]{ifsym}

\makeatother

\usepackage{babel}
\providecommand{\corollaryname}{Corollary}
\providecommand{\definitionname}{Definition}
\providecommand{\lemmaname}{Lemma}
\providecommand{\theoremname}{Theorem}

\begin{document}

\title{Throughput Analysis for Full-Duplex Wireless Networks with Imperfect Self-interference Cancellation}

\author{Zhen Tong\textsuperscript{1},~\IEEEmembership{Student Member,~IEEE,} and Martin Haenggi\textsuperscript{1},~\IEEEmembership{Fellow,~IEEE}}
\maketitle
\footnotetext[1]{Zhen Tong and Martin Haenggi are with Department of Electrical Engineering, 
University of Notre Dame, 
              Notre Dame, IN 46556, USA
              E-mail: {\{ztong1,mhaenggi\}@nd.edu\\
              Part of this work will be presented at WCNC 2015~\cite{Tong15WCNC}.
              }
 }

\begin{abstract}
This paper investigates the throughput for wireless network with full-duplex radios using stochastic geometry. Full-duplex (FD) radios can
exchange data simultaneously with each other. On the other hand, the
downside of FD transmission is that it will inevitably cause extra
interference to the network compared to half-duplex (HD) transmission. Moreover, the residual self-interference has negative effects on the network throughput.
In this paper, we focus on a wireless network of nodes with both HD
and FD capabilities and derive and optimize the throughput
in such a network. Our analytical result shows that if the network is adapting an ALOHA
protocol, the maximal throughput is achieved by scheduling all concurrently transmitting nodes
to work in either FD mode or HD mode depending on one simple condition. Moreover, the effects of imperfect self-interference cancellation on the signal-to-interference ratio (SIR) loss and throughput are also analyzed based on our mathematical model. We rigorously quantify the impact of imperfect self-interference cancellation on the throughput gain, transmission range, and other metrics, and we establish the minimum amount of self-interference suppression needed for FD to be beneficial.
\end{abstract}

\section{Introduction}

Traditionally, radio transceivers are subject to a  HD constraint
because of the crosstalk between the transmit and receive chains.
The self-interference caused by the transmitter at the receiver if using
FD transmission overwhelms the desired received signal from the partner
node since it is  much stronger than the desired received
signal. Therefore, current radios all use orthogonal signaling dimensions, i.e., time division
duplexing (TDD) or frequency division duplexing (FDD), to achieve bidirectional
communication. 

FD communication can potentially double the throughput if the self-interference
can be well mitigated. FD radios have been successfully implemented in the industrial, scientific and medical (ISM) radio bands in laboratory environments in the past few years~\cite{ChoJai10Mobicom, DuaSab10Asilomar, JaiCho11Mobicom, SahPat11X}.  Key
to the success are novel analog and digital self-interference cancellation
techniques and/or spatially separated transmit and receive antennas.
A FD system with only one antenna has also been implemented in \cite{Knox12WAMICON}
by using specially designed circulator and the FD WiFi radio with one antenna and one circulator has been prototyped in \cite{Bharadia13}. In general, the main idea of FD transmission
is to let the receive chain of a node remove the self-interference
caused by the known signal from its transmit chain, so that reception
can be concurrent with transmission. A novel signaling technique
was proposed in \cite{Guo10Allerton} to achieve virtual FD with applications
in neighbor discovery \cite{Guo2-13} and mutual broadcasting \cite{Guo13}  with its prototyping presented in \cite{Tong14RODD}.

From a theoretical perspective, the two-way transmission capacity
of wireless ad hoc networks has been studied in \cite{Vaze11} for
a FDD model. A FD cellular system has been analyzed in \cite{Goyal2013}
where the throughput gain has been illustrated via extensive simulation
for a cellular system with FD
base station and HD mobile users. The throughput gain of
single-cell multiple-input and multiple-output (MIMO) wireless systems with FD radios has been
quantified in \cite{Barghi12}. A capacity analysis of FD and HD transmissions with bounded
radio resources has been presented in \cite{Aggarwal12ITW} with focus only
on a single-link system. \cite{Ju12,Kim14} evaluate the capacity of FD ad hoc networks and alleviate the capacity
degradation due to the extra interference of FD by using beamforming and an ARQ protocol, respectively. Both
capacity analyses in \cite{Ju12,Kim14} are based on perfect self-interference cancellation and the approximation
that the distances of the two interfering nodes of a FD link to the desired receiver are the same.

In this paper, the impacts of FD transmission on the network throughput
are explored. On the one hand, FD transmission allows bidirectional communication
between two nodes simultaneously and therefore potentially doubles the throughput.
On the other hand, the extra interference caused by FD transmissions and imperfect self-interference cancellation can
degrade the throughput gain over HD, which makes it unclear whether FD
can actually outperform HD.
This paper utilizes the powerful analytical tools from stochastic geometry
to study the throughput performance of a wireless network of nodes with both FD and HD capabilities. Our results analytically
show that for an ALOHA MAC protocol, FD always outperforms HD in terms of throughput if perfect self-interference cancellation is assumed. However, for a path loss exponent $\alpha$, the achievable throughput gain is upper bounded by $\frac{2\alpha}{\alpha + 2}$, i.e., it ranges from  $0$-$33\%$ for the practical range $\alpha \in (2,4]$.
This result holds for arbitrary node densities, link distances and SIR regimes. Moreover, we model imperfect self-interference cancellation and quantify its effects on the throughput. Imperfect self-interference cancellation causes a SIR loss in the FD transmission and thus reduces the throughput gain between the FD network and HD network. Tight bounds on the SIR loss are obtained using the concept of horizontal shifts of the SIR distribution. The amount of self-interference cancellation determines if HD or FD is preferable in the networks.


\section{Network Model\label{sec:Network-Model}}

Consider an independently marked Poisson point process (PPP) \cite{Haenggi12book}
$\hat{\Phi}=\left\{ \left(x_{i},m(x_{i}),s(x_{i})\right)\right\} $
on $\mathbb{R}^{2}\times\mathbb{R}^{2}\times\left\{ 0,1,2\right\} $
where the ground process $\Phi=\left\{ x_{i}\right\} $ is a PPP with density $\lambda$
and $m(x_{i})$ and $s({x_{i}})$ are the marks of point
$x_{i}$. The mark $m(x_{i})$ is the location of the node that $x_{i}$
 communicates with. Here, we fix $\left\Vert x-m(x)\right\Vert =R$, $\forall x\in\Phi$,
i.e., $R$ is the distance of all links. Therefore, $m(x_{i})$ can also be written as $m(x_{i})=x_{i}+R\left(\cos\varphi_{i},\sin\varphi_{i}\right)$,
where the angles $\varphi_{i}$ are independent and uniformly distributed on $\left[0,2\pi\right]$. The link distance $R$ can also be random without affecting the main conclusions since we can always derive the results by first conditioning on $R$ and then averaging over $R$. We define $m(\Phi)=\{m(x): x\in\Phi\}$, which is also a PPP of density $\lambda$.
The mark $s(x_{i})$ indicates the independently chosen state of the link that consists
of $x_{i}$ and $m(x_{i})$: $s(x_{i})=0$ means the link is silent,
$s(x_{i})=1$ means the link is in HD mode, and $s(x_{i})=2$ means it is in FD mode. HD means that in a given
time slot the transmission is unidirectional, i.e., only from $x_{i}$
to $m(x_{i})$, while FD means that $x_{i}$ and $m(x_{i})$
are transmitting to each other concurrently. Therefore, for any link there are three
states: silence, HD, and FD. Assume that a link is in the state of
silence with probability $p_{0}$, HD with probability $p_{1}$ and
FD with probability $p_{2}$, where $p_{0}+p_{1}+p_{2}=1$. $p_{1}$
and $p_{2}$ are the medium access probabilities (MAPs) for HD and
FD modes, respectively. As a result, $\Phi=\bigcup_{i=0}^{2}\Phi_{[i]}$,
where $\Phi_{[i]}=\left\{ x\in\Phi: s(x)=i\right\} $ with density $\lambda p_{i}$
and $i\in\left\{ 0,1,2\right\}$. From the marking theorem \cite[Thm. 7.5]{Haenggi12book},
these three node sets $\Phi_{[i]}$ are independent. We call the link consisting of a node $x_{0}$ and its mark $m(x_{0})$ as the \textit{typical link}.

\begin{figure}[h]
\vspace{-3mm}
\begin{centering}
\includegraphics[width=\figwidth]{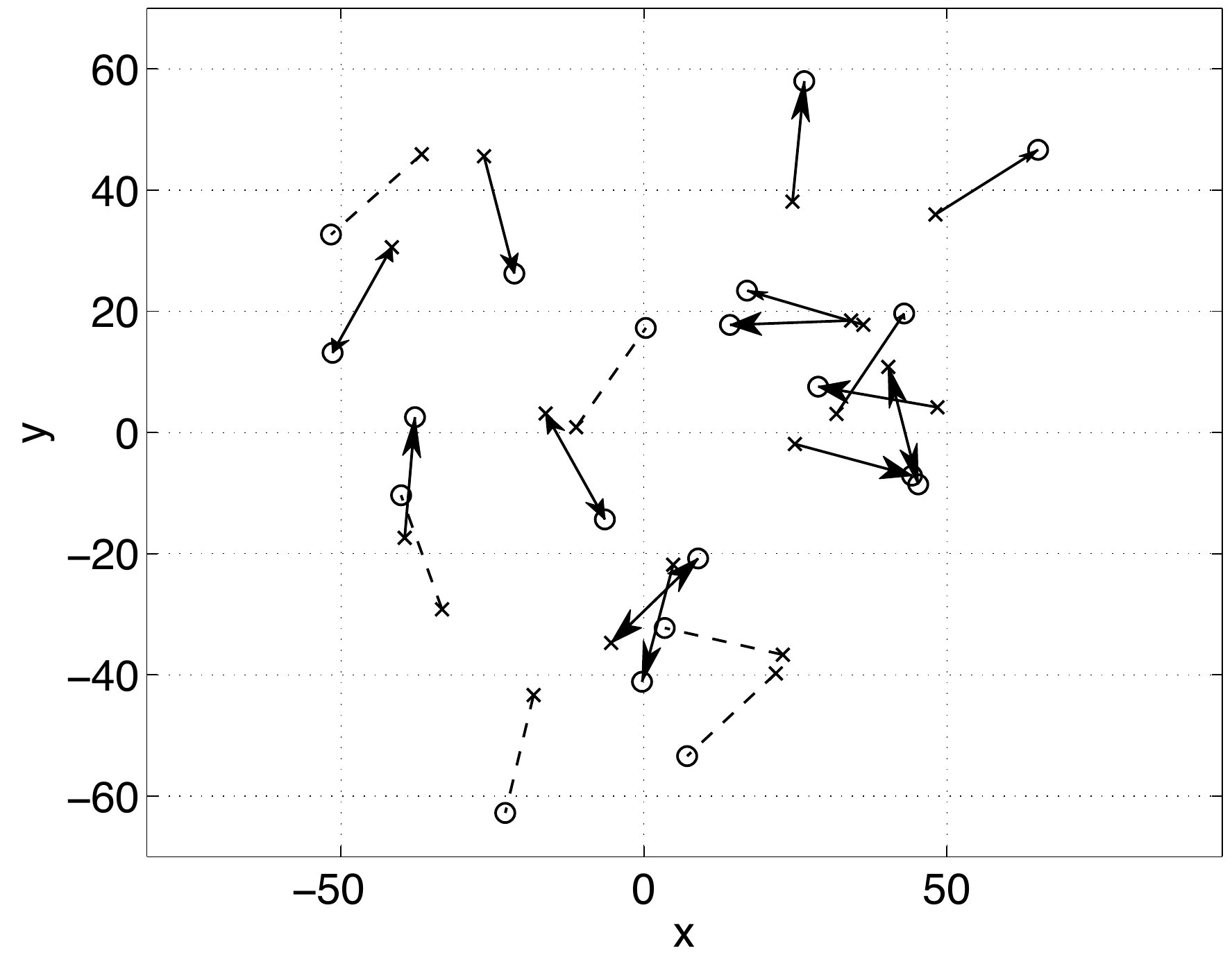}
\par\end{centering}
\caption{An example of the class of wireless networks considered in this paper. The dashed lines indicate the link is silent, the arrows
mean the link is in HD mode, and the double arrows in FD mode. The $\times$'s form $\Phi$ while the $\circ$'s form $m(\Phi)$.}
\label{Fig:0}
\end{figure}

The marked point process $\hat{\Phi}$ can be used to
model a wireless network of nodes with both FD and HD capabilities.
The self-interference in the FD links is assumed to be cancelled
imperfectly with residual self-interference-to-power ratio (SIPR) $\beta$, i.e., when the transmit power of a node is $P$, the residual self-interference is $\beta P$. The parameter $\beta$ quantifies the amount of self-interference cancellation, and $-10\log_{10} \beta$ is the self-interference cancellation in dB. When $\beta =0$, there is perfect self-interference cancellation, while for $\beta=1$, there is no self-interference cancellation. An example of a realization of such a wireless
network is illustrated in Figure \ref{Fig:0}. In the following, we will use this model to study the performance
of wireless networks with FD radios. 

In this network setup, we use the SIR model where a transmission
attempt from $x$ to $y$ is considered successful if 
\begin{equation}
\mbox{SIR}_{y}=\frac{P_{xy}Kh_{xy}l(x,y)}{\sum_{z\in\tilde{\Phi}\backslash\left\{ x\right\} }P_{yz}Kh_{yz}l(z,y)+ \beta P_{xy}\mathbbm{1}_{xy}^{\rm FD}}>\theta,\label{SIR}
\end{equation}
where $\tilde{\Phi}$ is the set of transmitting nodes in a given time slot, $\theta$ is the SIR threshold, $h_{xy}$ and $h_{zy}$ are the fading power coefficients with
mean $1$ from the desired transmitter $x$ and the interferer $z$
to $y$ respectively, and $\mathbbm{1}_{xy}^{\rm FD}$ is the indicator function that the link $xy$ is in FD mode. The inclusion of $\mathbbm{1}_{xy}^{\rm FD}$ means that the interference of FD links has an extra term due to the imperfect self-interference cancellation. The transmit powers $P_{xy} = P$ when link $xy$ is active. We focus on the Rayleigh fading case for both
the desired link and interferers. $K$ is a unitless constant that depends on the antenna characteristics and the average channel attenuations. $K=G_{\rm tx} G_{\rm rx}\left(\frac{c_{\rm L}}{4\pi f_c}\right)^2$, where $c_{\rm L}$ is the speed of light, $f_c$ is the carrier frequency, and $G_{\rm tx}$ and $G_{\rm rx}$ are the antenna gain at the transmitter and receiver, respectively. The path loss function $l(x,y)$
between node $x$ and $y$ is $l(x,y)=\left\Vert x-y\right\Vert ^{-\alpha}$,
where $\alpha>2$ is the path-loss exponent. If $y$ is at the origin, the index $y$ will be omitted, i.e., $l(x,\mathbf{0})\equiv l(x)$. Also, we
call a given set of system parameters $(\lambda,\theta,R,\alpha)$
a \textit{network configuration}. We will show that
some conclusions hold regardless of the network configuration.

\section{Success Probability\label{sec:Success-Probability}}

Our first metric of interest is the success probability, defined as
\begin{equation}
p_{s}\triangleq\mathbb{P}(\mbox{SIR}_{y}>\theta),\label{eq:ps}
\end{equation}
which is also the complementary cumulative distribution function (ccdf) of the SIR. Without changing the distribution of the point process, we may assume that
the receiver $y$ is at the origin. This implies there is a transmitter at fixed distance
$R$ from the origin. The success probability plays an
important role in determining the throughput, as will be described
in the following section.
\subsection{Derivation of the success probability and its bounds}
Before obtaining the unconditional success probability given in (\ref{eq:ps}), we first derive the conditional success probabilities given that a link is HD or FD\footnote{When the link is inactive, the conditional success probability is obviously zero by (\ref{SIR}).}. We denote the success probabilities conditioning that the typical link is HD and FD as $p_{s}^{\rm HD}$ and $p_{s}^{\rm FD}$, respectively. The following theorem gives the conditional success probabilities $p_{s}^{\rm HD}$ and $p_{s}^{\rm FD}$ of the FD/HD-mixed
wireless network modeled by the marked PPP:
\begin{thm}\label{sucProb}
In a wireless network described by the marked PPP $\hat{\Phi}$, the conditional success
probability $p_{s}^{\rm HD}$ is given by 
\begin{equation}
p_{s}^{\rm HD}=\exp(-\lambda p_{1}H(\theta R^{\alpha},\alpha))\exp(-\lambda p_{2}F(\theta R^{\alpha},\alpha,R)),\label{eq:ps-2}
\end{equation}
where $H(s,\alpha)\triangleq\frac{\pi^{2}\delta s^{\delta}}{\sin(\pi\delta)}$
with $\delta\triangleq2/\alpha$ and
\ifCLASSOPTIONonecolumn
\begin{equation}
F(s,\alpha,R)\triangleq\int_{0}^{\infty}\left(2\pi-\frac{1}{1+sr^{-\alpha}}\int_{0}^{2\pi}\frac{d\varphi}{1+s\left(r^{2}+R^{2}+2rR\cos\varphi\right)^{-\alpha/2}}\right)rdr,\label{F}
\end{equation}
\else
\begin{equation}F(s,\alpha,R)\triangleq \int_{0}^{\infty}\left(2\pi-\frac{1}{1+sr^{-\alpha}}K(s,r,R,\alpha)\right)rdr \label{F}\end{equation}
with $K(s,r,R,\alpha)\triangleq\int_{0}^{2\pi}\frac{d\varphi}{1+s\left(r^{2}+R^{2}+2rR\cos\varphi\right)^{-\alpha/2}},$
\fi
and the conditional success
probability $p_{s}^{\rm FD}$ is given by 
\begin{equation}
p_{s}^{\rm FD}=\kappa p_{s}^{\rm HD},\label{eq:ps-3}
\end{equation}
where $\kappa \triangleq e^{-\frac{\theta R^{\alpha}\beta}{K}}$.
\end{thm}
\begin{IEEEproof}
Conditional on that the link is active, the SIR from (\ref{SIR}) can be rewritten as
\begin{equation}
\mbox{SIR}_{y}=\frac{h_{xy}l(x,y)}{\sum_{z\in\tilde{\Phi}\backslash\left\{ x\right\} }h_{yz}l(z,y)+ \frac{\beta \mathbbm{1}_{xy}^{\rm FD}}{K}}>\theta,\label{SIR2}
\end{equation}
by dividing both numerator and denumerator by $PK$.
As a result, it is equivalent to a network where each node transmit with unit power while the SIPR $\beta$ is scaled by $K$. Hence, with Rayleigh fading, the desired signal strength $S$ at the
receiver at the origin is exponential, i.e., $S=hR^{-\alpha}$. Conditional on that the link is HD, the interference $I$ consists of two parts: the
interference from the HD nodes $\Phi_{[1]}$ and the interference from the FD nodes $\Phi_{[2]}$. Hence, it can be expressed as:
\[
I=\sum_{x\in\Phi_{[1]}}h_{x}l(x)+\sum_{x\in\Phi_{[2]}}\left(h_{x}l(x)+h_{m(x)}l(m(x))\right).
\]
The Laplace transform of the interference follows as
\ifCLASSOPTIONonecolumn
\begin{eqnarray}
L_{I}(s) & = & \mathbb{E}e^{-s\left(\sum_{x\in\Phi_{[1]}}h_{x}l(x)+\sum_{x\in\Phi_{[2]}}\left(h_{x}l(x)+h_{m(x)}l(m(x))\right)\right)}\nonumber \\
 & = & \mathbb{E}\left(\prod_{x\in\Phi_{[1]}}e^{-sh_{x}l(x)}\prod_{x\in\Phi_{[2]}}e^{-s\left(h_{x}l(x)+h_{m(x)}l(m(x))\right)}\right)\nonumber \\
 & \overset{\left(a\right)}{=} & \mathbb{E}\left(\prod_{x\in\Phi_{[1]}}e^{-sh_{x}l(x)}\right)\mathbb{E}\left(\prod_{x\in\Phi_{[2]}}e^{-s\left(h_{x}l(x)+h_{m(x)}l(m(x))\right)}\right),\label{eq:Two}
\end{eqnarray}
\else
\begin{align}
L_{I}(s)  
 & =  \mathbb{E}\left(\prod_{x\in\Phi_{\left[1\right]}}e^{-sh_{x}l(x)} \prod_{x\in\Phi_{\left[2\right]}}e^{-s\left(h_{x}l(x)+h_{m(x)}l(m(x))\right)}\right)\nonumber \\
 &  \overset{\left(a\right)}{=}  \mathbb{E}\left(\prod_{x\in\Phi_{\left[1\right]}}e^{-sh_{x}l(x)}\right)\cdot\nonumber\\
 & \qquad \mathbb{E}\left(\prod_{x\in\Phi_{\left[2\right]}}e^{-s\left(h_{x}l(x)+h_{m(x)}l(m(x))\right)}\right),\label{eq:Two}
\end{align}
\fi
where (a) follows from the fact that $\Phi_{[1]}$ and
$\Phi_{\left[2\right]}$ are independent PPPs from the marking theorem
 \cite[Thm. 7.5]{Haenggi12book}. The first term in the
product of (\ref{eq:Two}) is the Laplace transform of the interference
of the PPP $\Phi_{[1]}$, given by \cite[page 103]{Haenggi12book}:
\begin{align*}
L_{I_{1}}(s) & =  \mathbb{E}\left(\prod_{x\in\Phi_{\left[1\right]}}e^{-sh_{x}l(x)}\right)\\
 & =  \exp(-\lambda p_{1}H(s,\alpha)).
\end{align*}

The second term in the product of (\ref{eq:Two}) can be written as
follows:\ifCLASSOPTIONonecolumn
\begin{align}
L_{I_{2}}(s) & =  \mathbb{E}\left(\prod_{x\in\Phi_{[2]}}e^{-s\left(h_{x}l(x)+h_{m(x)}l(m(x))\right)}\right)\nonumber \\
 & =  \mathbb{E}\left(\prod_{x\in\Phi_{[2]}}\frac{1}{1+sl(x)}\frac{1}{1+sl(m(x))}\right)\label{eq:I2}\\
 & \overset{\left(a\right)}{=}  \exp\left(-\lambda p_{2}\int_{\mathbb{R}^{2}}\left(1-\frac{1}{1+sl(x)}\frac{1}{1+sl(m(x))}\right)dx\right)\label{eq:I2-1}\\
 & =  \exp(-\lambda p_{2}F(s,\alpha,R)),\label{L2}
\end{align}
where (a) follows from the probability generating functional of the
PPP.
\else
\begin{align}
L_{I_{2}}(s) & =  \mathbb{E}\left(\prod_{x\in\Phi_{[2]}}e^{-s\left(h_{x}l(x)+h_{m(x)}l(m(x))\right)}\right)\nonumber\\
 & =  \mathbb{E}\left(\prod_{x\in\Phi_{[2]}}\frac{1}{1+sl(x)}\frac{1}{1+sl(m(x))}\right)\label{eq:I2}\\
 & \overset{\left(a\right)}{=}  \exp\left(-\lambda p_{2}\int_{\mathbb{R}^{2}}\left(1-v(x)\right)\right)\label{eq:I2-1}\\
 & =  \exp(-\lambda p_{2}F(s,\alpha,R)),\label{L2}
\end{align}
where (a) follows from the probability generating functional of the
PPP with $v(x)=\frac{1}{1+sl(x)}\frac{1}{1+sl(m(x))}.$
\fi
As a result, the success probability is 
\begin{align}
p_{s}^{\rm HD} & =  L_{I_{1}}(\theta R^{\alpha})L_{I_{2}}(\theta R^{\alpha})\label{HDps}\\
 & =  \exp(-\lambda p_{1}H(\theta R^{\alpha},\alpha))\exp(-\lambda p_{2}F(\theta R^{\alpha},\alpha,R)),
\end{align}
which completes the proof of $p_{s}^{\rm HD}$.

Conditional on a FD link, there is an extra term in the interference, which is the residual self-interference scaled by the constant $K$. Hence, the interference for a FD link consists of three parts as follows:
\[
I=\sum_{x\in\Phi_{\left[1\right]}}h_{x}l(x)+\sum_{x\in\Phi_{\left[2\right]}}\left(h_{x}l(x)+h_{m(x)}l(m(x))\right) + \frac{\beta}{K}.
\]
The first two terms are the same as in the proof of $p_s^{\rm HD}$ while the third term is the residual self-interference. Hence, the Laplace transform of the interference follows as
\begin{eqnarray}
L_{I}(s) & = & L_{I_{1}}(s) L_{I_{2}}(s) e^{-\frac{s\beta}{K}}.\label{eq:Two2}
\end{eqnarray}

As a result, the conditional success probability $p_{s}^{\rm FD}$ is 
\begin{align*} 
p_{s}^{\rm FD} & =  L_{I_{1}}(\theta R^{\alpha})L_{I_{2}}(\theta R^{\alpha})e^{-\frac{\theta R^{\alpha}\beta}{K}}\\
 & = \kappa p_{s}^{\rm HD},
\end{align*}
where the last step is from (\ref{HDps}).
\end{IEEEproof}
Alternatively, $p_s^{\rm HD}$ can also be derived using the results for the  Gauss-Poisson process\cite{Guo14ISIT}.

The fact that the conditional success probability $p_s^{\rm HD}$ (and the Laplace transform of the interference)
is a product of two terms follows from the
independence of the point processes $\Phi_{\left[i\right]}$. The names of the functions $H$ and $F$ are chosen to reflect the fact that they represent the case of half- and full-duplex, respectively.

The residual self-interference for FD links simply adds an exponential factor to the success probability for HD links, which is similar to the effect of noise as in \cite[page 105]{Haenggi12book}. Theorem \ref{sucProb} also reveals the connection between two conditional success probability  $p_{s}^{\rm FD}$ and  $p_{s}^{\rm HD}$. As expected,  $p_{s}^{\rm FD}\leq p_{s}^{\rm HD}$, with equality for perfect self-interference cancellation.

The unconditional success probability can be easily obtained from the results in Theorem \ref{sucProb}.
\begin{cor}\label{sucProb_FD2}
In a wireless network described by the marked PPP $\hat{\Phi}$, the unconditional success
probability $p_{s}$ is given by 
\begin{equation}
p_{s}=\left(p_1+\kappa p_2 \right) e^{-\lambda p_{1}H(\theta R^{\alpha},\alpha)}e^{-\lambda p_{2}F(\theta R^{\alpha},\alpha,R)}.\label{eq:ps-4}
\end{equation}
\end{cor}
\begin{IEEEproof}
Since a link is HD with probability $p_1$ and FD with probability $p_2$, the unconditional success probability from (\ref{eq:ps}) is the average
\begin{equation}
p_s = p_1 p_{s}^{\rm HD} + p_2 p_{s}^{\rm FD}.\label{ps0}
\end{equation}
Inserting the results from (\ref{eq:ps-2}) and (\ref{eq:ps-3}), we have (\ref{eq:ps-4}).
\end{IEEEproof}

The (un)conditional success probabilities are not in strict closed-form due to the integral form of $F(\theta R^{\alpha},\alpha,R)$. However,
tight simple bounds can be obtained.
\begin{thm}
The conditional success probability $p_{s}^{\rm HD}$ is lower and upper bounded by
\begin{equation}
\underline{p}_{s}=\exp(-\lambda(p_{1}+2p_{2})H(\theta R^{\alpha},\alpha))\label{lb}
\end{equation}
and
\begin{equation}
\overline{p}_{s}=\exp(-\lambda(p_{1}+p_{2}(1+\delta))H(\theta R^{\alpha},\alpha)).\label{ub}
\end{equation}
and, similarly, $p_{s}^{\rm FD}$ is bounded as 
 \begin{equation} \kappa \underline{p}_{s} \leq p_{s}^{\rm FD} \leq \kappa \overline{p}_{s}.\label{b1}\end{equation}The unconditional success probability is lower and upper bounded as
 \begin{equation}
 (p_1 + \kappa p_2)\underline{p}_{s} \leq p_s \leq (p_1 + \kappa p_2)\overline{p}_{s}.\label{b2}
 \end{equation}
\end{thm}
\begin{IEEEproof}
Bounds only need to be established for the second term of the product in the conditional success
probability that contains the integral $F(\theta R^{\alpha},\alpha,R)$.

Lower Bound: From (\ref{eq:I2-1}),
\ifCLASSOPTIONonecolumn
\begin{eqnarray}
L_{I_{2}}(s) & = & \mathbb{E}\left(\prod_{x\in\Phi_{\left[2\right]}}e^{-s\left(h_{x}l(x)+h_{m(x)}l(m(x))\right)}\right)\nonumber \\
 & \overset{\left(a\right)}{\geq} & \mathbb{E}\left(\prod_{x\in\Phi_{\left[2\right]}}e^{-sh_{x}l(x)}\right)\mathbb{E}\left(\prod_{x\in\Phi_{\left[2\right]}}e^{-sh_{m(x)}l(m(x))}\right)\label{eq:lb2}\\
 & \overset{\left(b\right)}{=} & \exp(-2\lambda p_{2}H(s,\alpha)),\label{K1}
\end{eqnarray}
\else
\begin{align}
L_{I_{2}}(s) & =  \mathbb{E}\left(\prod_{x\in\Phi_{\left[2\right]}}e^{-s\left(h_{x}l(x)+h_{m(x)}l(m(x))\right)}\right)\nonumber \\
 & \overset{\left(a\right)}{\geq}  \mathbb{E}\left(\prod_{x\in\Phi_{\left[2\right]}}e^{-sh_{x}l(x)}\right)\mathbb{E}\left(\prod_{x\in\Phi_{\left[2\right]}}e^{-sh_{m(x)}l(m(x))}\right)\label{eq:lb2}\\
 & \overset{\left(b\right)}{=} \exp(-2\lambda p_{2}H(s,\alpha)),\label{K1} 
\end{align}
\fi 
where (a) follows from the FKG inequality \cite[Thm 10.13]{Haenggi12book} since
both $\prod_{x\in\Phi}e^{-sh_{x}l(x)}$ and \ifCLASSOPTIONonecolumn \\ \else \fi $\prod_{x\in\Phi}e^{-sh_{m(x)}l(m(x))}$
are decreasing random variables. In (\ref{eq:lb2}), the first term
is similar to the calculation of $L_{I_{1}}(s)$ with $\Phi_{\left[1\right]}$
replaced by $\Phi_{\left[2\right]}$ while in the second term, $m(\Phi_{\left[2\right]}) $ is a PPP with the same density as $\Phi_{\left[2\right]}$
due to the displacement theorem \cite[page 35]{Haenggi12book}. As a result, the two factors in (\ref{eq:lb2}) are equal, and
\begin{align*}
p_{s}^{\rm HD} 
 & \geq  L_{I_{1}}(\theta R^{\alpha})\exp(-2\lambda p_{2}H(\theta R^{\alpha},\alpha))
 = \underline{p}_{s}.
\end{align*}

Upper Bound: From (\ref{eq:I2}),
\begin{align*}
L_{I_{2}}(s) & =  \mathbb{E}\left(\prod_{x\in\Phi_{\left[2\right]}}\frac{1}{1+sl(x)}\frac{1}{1+sl(m(x))}\right)\\
 & \leq  \left\{ K_{1}(s,\alpha)K_{2}(s,\alpha)\right\} ^{\frac{1}{2}},
\end{align*}
which follows from the Cauchy-Schwarz inequality with $K_{1}(s,\alpha)=\mathbb{E}\left(\prod_{x\in\Phi_{\left[2\right]}}\frac{1}{\left(1+sl(x)\right)^{2}}\right)$
and $K_{2}(s,\alpha)=\mathbb{E}\left(\prod_{x\in\Phi_{\left[2\right]}}\frac{1}{\left(1+sl(m(x))\right)^{2}}\right)$. We have
\begin{align*}
K_{1}(s,\alpha)
 & =  \exp\left(-2\pi\lambda p_{2}\int_{0}^{\infty}\left(1-\frac{1}{\left(1+sr^{-\alpha}\right)^{2}}\right)rdr\right)\\
 & =  \exp\left(-\pi\lambda p_{2}(1+\delta)\Gamma(1+\delta)\Gamma(1-\delta)s^{\delta}\right)\\
 & =  \exp(-\lambda p_{2}(1+\delta)H(s,\alpha)),
\end{align*}where $\Gamma(\cdot)$ is the gamma function.
$K_{2}(s,\alpha)=K_{1}(s,\alpha)$ because $m(\Phi_{\left[2\right]})$
is a PPP with the same density as $\Phi_{\left[2\right]}$. As
a result, 
\begin{align}
L_{I_{2}}(s) & \leq  \left\{ K_{1}(s,\alpha)K_{2}(s,\alpha)\right\} ^{\frac{1}{2}}\\
 & =  \exp(-\lambda p_{2}(1+\delta)H(s,\alpha)).\label{K2}
\end{align}
Therefore, 
\begin{align*}
p_{s}^{\rm HD} 
 & \leq  e^{-\lambda p_{1}H(\theta R^{\alpha},\alpha)}e^{-\lambda p_{2}(1+\delta)H(\theta R^{\alpha},\alpha)}
 = \overline{p}_{s}.
\end{align*}
The lower and upper bounds of $p_{s}^{\rm FD}$ and $p_{s}$ simply follow from (\ref{eq:ps-3}) and (\ref{eq:ps-4}). 
\end{IEEEproof}
The lower bound can be intuitively
understood as lower bounding the interference of the FD nodes (which
are formed by two {\em dependent} PPPs) by that of two {\em independent} PPPs with the same
density.

The upper bound turns out to be the same as the result obtained
by assuming $l(x)=l(m(x))$, $\forall x \in \Phi$, i.e., the distances between the receiver
at the origin and the interfering pair from the FD links are the same. 
Indeed, assuming $l(x)=l(m(x))$, we have\begin{align*}
\tilde{L}_{I_{2}}(s) & =  \mathbb{E}\left(\prod_{x\in\Phi_{\left[2\right]}}e^{-s\left(h_{x}+h_{m(x)}\right)l(x)}\right)\\
 & =  \exp\left(-\pi\lambda p_{2}\mathbb{E}\left[\left(h_{x}+h_{m(x)}\right)^{\delta}\right]\Gamma\left(1-\delta\right)s^{\delta}\right)\\
 & =  \exp(-\pi\lambda p_{2}\Gamma(2+\delta)\Gamma(1-\delta)s^{\delta})\\
 & =  \exp(-\pi\lambda p_{2}\left(1+\delta\right)\Gamma(1+\delta)\Gamma(1-\delta)s^{\delta})\\
 & =  \exp(-\lambda p_{2}(1+\delta)H(s,\alpha))
\end{align*}
where $\mathbb{E}\left[\left(h_{x}+h_{y}\right)^{\delta}\right]=\Gamma(2+\delta)$
since $h_{x}+h_{y}$ has an Erlang distribution
and $\Gamma(2+\delta)=(1+\delta)\Gamma(1+\delta)$. Hence, the approximated
success probability assuming $l(x)=l(m(x))$ is $\tilde{p}_{s} = L_{I_{1}}(\theta R^{\alpha})\tilde{L}_{I_{2}}(\theta R^{\alpha})= \overline{p}_{s}$. This result is not surprising. The equality holds for the Cauchy-Schwarz
inequality if $\prod_{x\in\Phi_{[2]}}\frac{1}{(1+sl(x))^{2}}$
and $\prod_{x\in\Phi_{[2]}}\frac{1}{(1+sl(m(x)))^{2}}$
are linearly dependent. Obviously, $l(x)=l(m(x))$ satisfies this condition. 
Therefore, we have $\tilde{p}_{s}=\overline{p}_{s}$
as expected. 

The horizontal gap between two success probability curves (or SIR distributions) is often quite insensitive to the success probability where it is evaluated and the path loss models, as pointed out in \cite{Haenggi14, Guo15TC}. The horizontal gap is defined as 
\begin{equation}
G(p) \triangleq \frac{p_{s_1}^{-1}(p)}{p_{s_2}^{-1}(p)}, \quad p\in (0,1),\label{hg}
\end{equation}
where $p_{s}^{-1}(p)$ is the inverse of the success probability and $p$ is the target success probability.
The sharpness of the upper and lower bounds of the success probabilities is established by the following corollary.
\begin{cor}
\label{cor:HGain} The horizontal gap between the upper and lower bound of the conditional success probability $p_s^{\rm HD}$ does not depend on the target success probability and is given by
\begin{equation}
G = \left(\frac{p_1+2p_2 }{p_1+p_2(1+\delta)}\right)^{{1}/{\delta}}.\label{HG}
\end{equation}
Furthermore, the horizontal gap between the lower and upper bound of the conditional success probability $p_s^{\rm FD}$ and the bounds of the unconditional success probability $p_s$ under perfect self-interference cancellation is also $G$. 
 \end{cor}
\begin{IEEEproof}
The horizontal gap can be obtained by setting $\overline{p}_{s}(\overline{\theta})= \underline{p}_{s}(\underline{\theta})$ and calculating $G = {\overline{\theta}}/{\underline{\theta}}$. From (\ref{lb}) and (\ref{ub}), we have
\ifCLASSOPTIONonecolumn
\begin{equation}
\exp(-\lambda(p_{1}+2p_{2})H(\overline{\theta} R^{\alpha},\alpha))= \exp(-\lambda(p_{1}+p_{2}(1+\delta))H(\underline{\theta}R^{\alpha},\alpha)).
\end{equation}
\else
\begin{equation}
e^{-\lambda(p_{1}+2p_{2})H(\overline{\theta} R^{\alpha},\alpha)}= e^{-\lambda(p_{1}+p_{2}(1+\delta))H(\underline{\theta}R^{\alpha},\alpha)}.
\end{equation}
\fi  
Solving the above equation for the ratio ${\overline{\theta}}/{\underline{\theta}}$, we obtain (\ref{HG}). When the self-interference cancellation is perfect, $\kappa = 1$. From (\ref{b1}) and (\ref{b2}), we obtain the same horizontal gap $G$ for $p_s^{\rm FD}$ and $p_s$. 
\end{IEEEproof}
\begin{figure}[h]
\begin{centering}
\includegraphics[width=\figwidth]{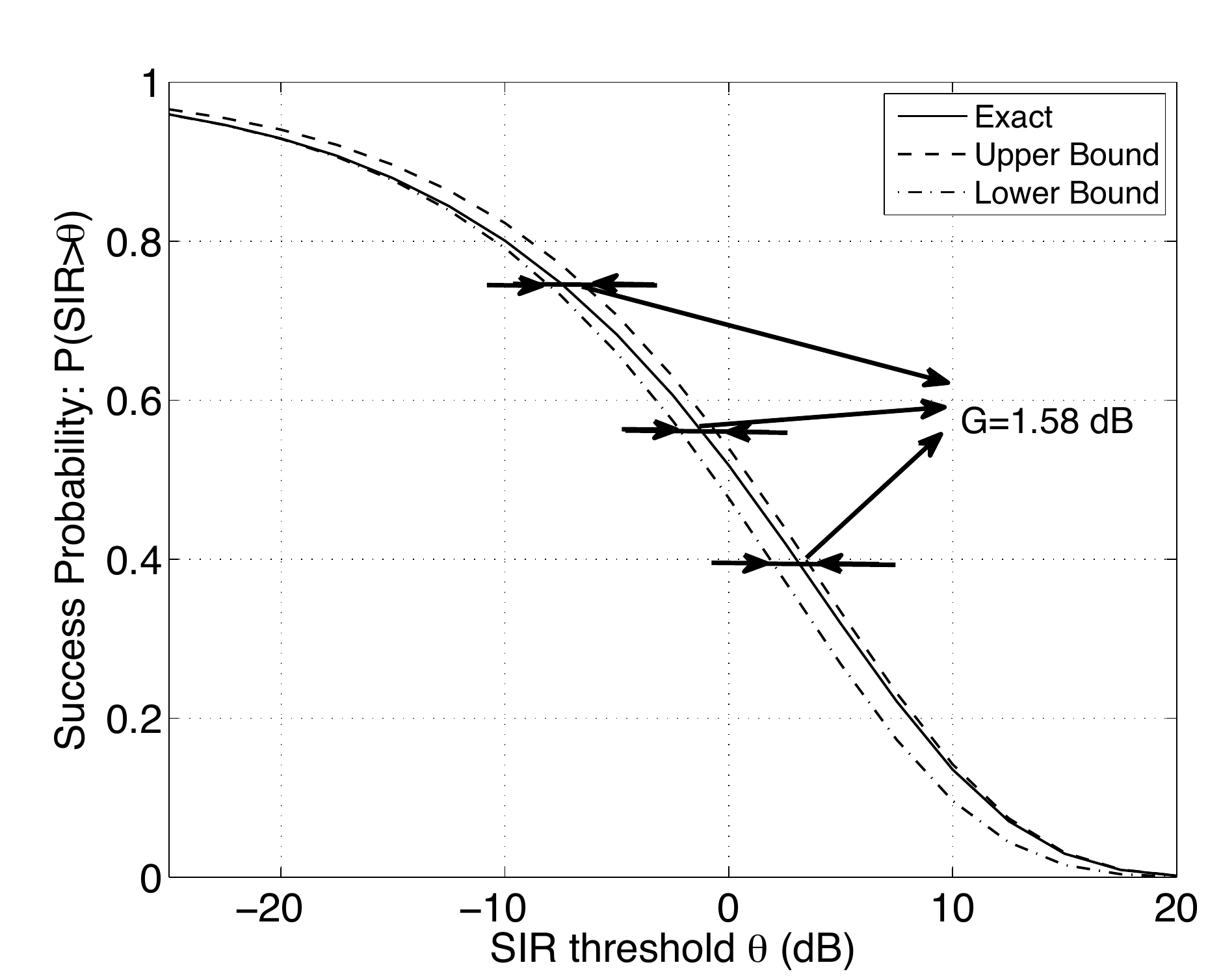}
\par\end{centering}
\caption{Comparison of unconditional success probability between the theoretical result from (\ref{eq:ps-4}) and its bounds
as a function of the SIR threshold $\theta$ in dB: $\alpha=4$, $\lambda=0.1$, $R=1$, 
$p_{0}=0$, $p_{1}=p_{2}=0.5$, $\beta = 0$. The horizontal gap from (\ref{HG}) is $G=36/25$ or $1.58$ dB for these parameters.}
\label{Fig:1}
\end{figure}

 From the above corollary, it is apparent that the gap $G$ is independent of the SIR threshold $\theta$ and the target success probability; it only depends on $\delta$ and the transmit probabilities. Also, for $\alpha \downarrow 2$, $G \downarrow 1$. Figure \ref{Fig:1} plots the unconditional success probability from (\ref{eq:ps-4})
and its closed-form upper and lower bounds under perfect self-interference cancellation as a function of the SIR threshold in dB. To obtain the exact curve, the double integral $F(\theta R^{\alpha},\alpha,R)$ is numerically evaluated. 
As seen, both bounds are tight with constant horizontal gap $G=1.58$ dB everywhere for $p_1=p_2 = 0.5$ and $\delta = 2/\alpha = 1/2$.

Furthermore, we can examine the relationship between the two key functions ${F(\theta R^{\alpha},\alpha,R)}$ and ${H(\theta R^{\alpha},\alpha)}$. The following corollary bounds their ratio:
\begin{cor}
\label{cor:F}The ratio ${F}/{H}$ is bounded as follows:
\begin{equation}
1+\delta \leq \frac{F(\theta R^{\alpha},\alpha,R)}{H(\theta R^{\alpha},\alpha)}\leq 2, \quad\forall\;\theta>0, R>0, \alpha>2.\label{F2}
\end{equation}
Moreover, the ratio is independent of the link distance $R$ and 
\begin{equation}
\lim_{\theta \rightarrow \infty}\frac{F(\theta R^{\alpha},\alpha,R)}{H(\theta R^{\alpha},\alpha)} = {1+\delta}.\label{FH2}
\end{equation}
 \end{cor}
\begin{IEEEproof}
From the proof of the upper and lower bounds of the conditional success probability $p_s^{\rm HD}$, i.e., (\ref{K1}) and (\ref{K2}),
we have
\begin{equation}
\exp(-2\lambda p_{2}H(s,\alpha)) \leq L_{I_{2}}(s) \leq  \exp(-\lambda p_{2}(1+\delta)H(s,\alpha)),
\end{equation}
where $L_{I_{2}}(s)=\exp(-\lambda p_{2}F(s,\alpha,R))$ from (\ref{L2}).
By taking the logarithm on both sides of the above, we have
\begin{equation}
-2\lambda p_{2}H(s,\alpha)\leq -\lambda p_{2}F(s,\alpha,R)  \leq  -\lambda p_{2}(1+\delta)H(s,\alpha),
\end{equation}
which leads to (\ref{F2}).

For the independence on the link distance $R$, since $H(\theta R^{\alpha},\alpha)=\frac{\pi^{2}\delta {\theta}^{\delta}R^2}{\sin\left(\pi\delta\right)}$, we need to prove that $F(\theta R^{\alpha},\alpha,R)$ is also proportional to $R^2$. By the change of the variable $r_1 = {r}/{R}$, we can express (\ref{F}) as
\ifCLASSOPTIONonecolumn
\begin{align}
F(\theta R^{\alpha},\alpha,R)
&= R^2\int_{0}^{\infty}\left(2\pi-\frac{1}{1+\theta r_{1}^{-\alpha}}\int_{0}^{2\pi}\frac{d\varphi}{1+\theta \left(r_{1}^{2}+1+2r_1\cos\varphi\right)^{-\alpha/2}}\right) r_1dr_1,\label{F_R}
\end{align}
\else
\begin{align}
F(\theta R^{\alpha},\alpha,R)
&= R^2\int_{0}^{\infty}\left(2\pi-\frac{K(\theta, r_1, 1, \alpha)}{1+\theta r_{1}^{-\alpha}}\right) r_1dr_1,\label{F_R}
\end{align}
\fi
which completes the proof of independence of link distance $R$. For the limit, by the change of the variable $r_2 = r_1 \theta^{-\frac{1}{\alpha}}$, we have
\ifCLASSOPTIONonecolumn
\begin{align}
F(\theta R^{\alpha},\alpha,R)
&= \theta^{\delta}R^2\int_{0}^{\infty}\left(2\pi-\frac{1}{1+ r_{2}^{-\alpha}}\int_{0}^{2\pi}\frac{d\varphi}{1+ \left(r_{2}^{2}+\theta^{-\delta}+2r_2\theta^{-\delta/2}\cos\varphi\right)^{-\alpha/2}}\right) r_2dr_2.\label{F_R2}
\end{align}
\else
\begin{align}
F(\theta R^{\alpha},\alpha,R)
&= \theta^{\delta}R^2\int_{0}^{\infty}\left(2\pi-\frac{K(1, r_2, \theta^{-\delta/2}, \alpha)}{1+ r_{2}^{-\alpha}}\right) r_2dr_2.\label{F_R2}
\end{align}
\fi
Therefore, 
\ifCLASSOPTIONonecolumn
\begin{align}
\lim_{\theta \rightarrow \infty}\frac{F(\theta R^{\alpha},\alpha,R)}{H(\theta R^{\alpha},\alpha)} &= \lim_{\theta \rightarrow \infty}\frac{\int_{0}^{\infty}\left(2\pi-\frac{1}{1+ r_{2}^{-\alpha}}\int_{0}^{2\pi}\frac{d\varphi}{1+ \left(r_{2}^{2}+\theta^{-\frac{2}{\alpha}}+2r_2\theta^{-\frac{1}{\alpha}}\cos\varphi\right)^{-\alpha/2}}\right) r_2dr_2}{\frac{\pi^{2}\delta }{\sin(\pi\delta)}}\\
&= \frac{\int_{0}^{\infty}\left(2\pi-\frac{2\pi}{\left(1+ r_{2}^{-\alpha}\right)^2}\right) r_2dr_2}{\frac{\pi^{2}\delta }{\sin(\pi\delta)}}\\
&=1+\delta.
\end{align}
\else
\begin{align}
\lim_{\theta \rightarrow \infty}\frac{F(\theta R^{\alpha},\alpha,R)}{H(\theta R^{\alpha},\alpha)} &= \lim_{\theta \rightarrow \infty}\frac{\int_{0}^{\infty}\left(2\pi-\frac{K(1, r_2, \theta^{-\delta/2}, \alpha)}{1+ r_{2}^{-\alpha}}\right) r_2dr_2}{\frac{\pi^{2}\delta }{\sin(\pi\delta)}}\\
&= \frac{\int_{0}^{\infty}\left(2\pi-\frac{2\pi}{\left(1+ r_{2}^{-\alpha}\right)^2}\right) r_2dr_2}{\frac{\pi^{2}\delta }{\sin(\pi\delta)}}\\
&=1+\delta.
\end{align}
\fi

\end{IEEEproof}
This corollary is useful in calculating the SIR loss, the maximal throughput, and
their bounds in the following. Also, we can conclude that the upper bound of the success probability is asymptotically exact as $\theta\to \infty$. It is also illustrated by \figref{Fig:1}.

\subsection{SIR loss due to FD operation}
In this subsection, we investigate the SIR loss caused by the FD operation in wireless networks described by $\hat{\Phi}$. Consider two extreme cases: one is the case where all concurrently
transmitting nodes work in HD mode, i.e., $p_1=1$, and the other
is where all concurrently transmitting nodes work in FD mode, i.e., $p_2 = 1$. The success probabilities of HD-only and FD-only networks follows from (\ref{eq:ps-2}) as
\begin{equation}
p_{s, \;p_1=1} = \exp(-\lambda H(\theta R^{\alpha},\alpha))\label{ps_HD}
\end{equation}

and \begin{equation}
p_{s, \;p_2=1} = \kappa\exp(-\lambda F(\theta R^{\alpha},\alpha,R)).\label{ps_FD}\end{equation}

\begin{figure}[h]
\begin{centering}
\includegraphics[width=\figwidth]{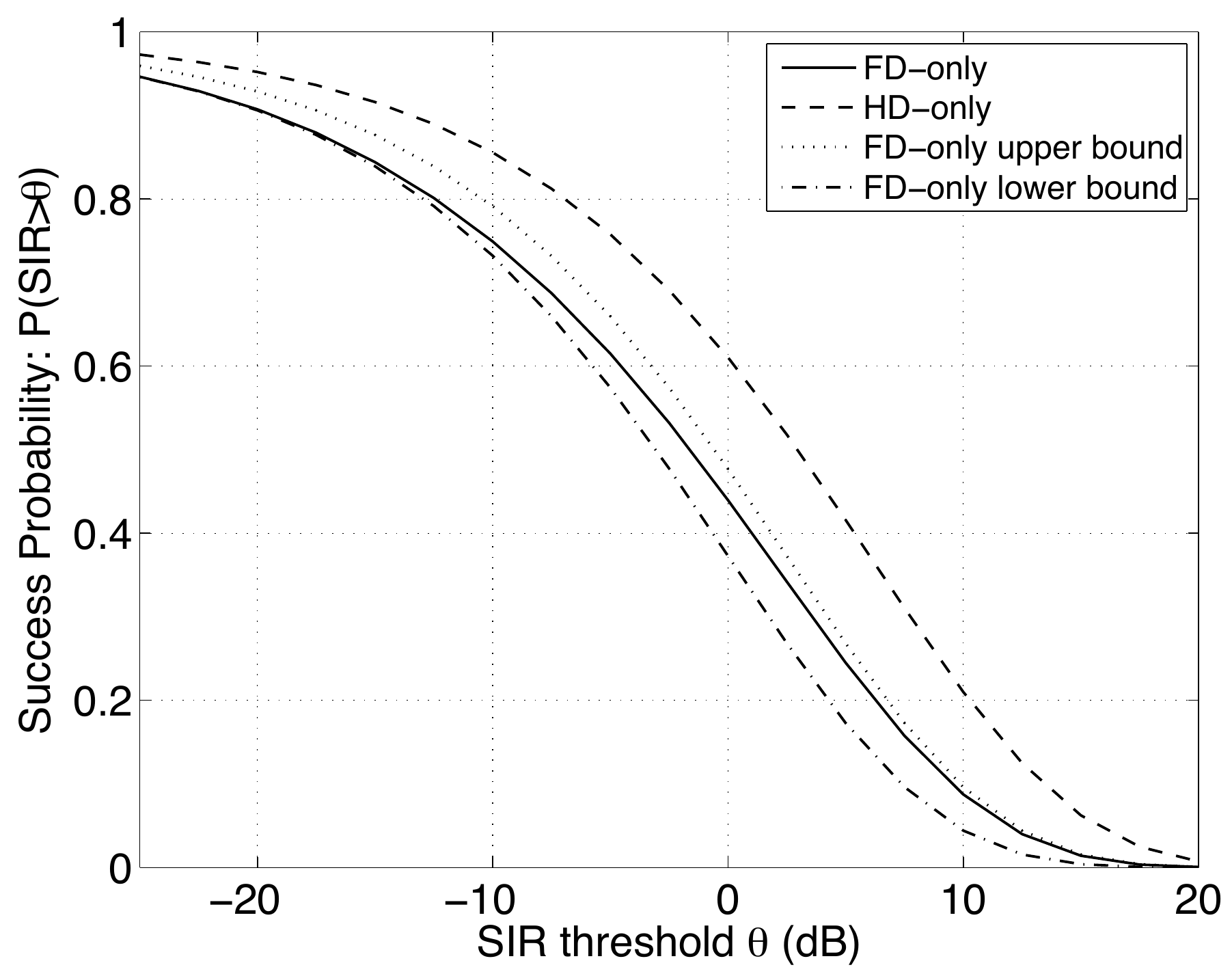}
\par\end{centering}
\caption{Comparison of success probabilities of FD-only networks, its bounds, and HD-only networks
as a function of the SIR threshold $\theta$ in dB under perfect self-interference cancellation: $\alpha=4$, $\lambda=0.1$, $R=1$, 
$p_{0}=0$, $\beta = 0$.}
\label{Fig:1-1}
\end{figure}

Figure \ref{Fig:1-1} plots the success probability of FD-only networks
and its upper and lower bounds as well as success probability of HD-only networks as a function of the SIR threshold in dB. Clearly, FD transmission in a FD-only wireless network leads to a SIR loss compared to its counterpart HD-only wireless network in the success probability, which is the ccdf of SIR. The SIR loss can be defined as the horizontal gap between two SIR distributions as follows from (\ref{hg}):
\begin{definition} The SIR loss between FD-only and HD-only networks is defined as
\begin{equation}
G(p) \triangleq \frac{\theta_{\rm HD}(p)}{\theta_{\rm FD}(p)}=\frac{p_{s, \;p_1=1}^{-1}(p)}{p_{s, \;p_2=1}^{-1}(p)},
\end{equation}
where $p$ is the target success probability and $p_{s}^{-1}$ is the inverse of the ccdf of the SIR. $\theta_{\rm HD}(p)$ is the SIR threshold when the target success probability is $p$, i.e., $\theta_{\rm HD}(p)=p_{s, \;p_1=1}^{-1}(p)$. Similarly, $\theta_{\rm FD}(p)=p_{s, \;p_2=1}^{-1}(p)$.
\end{definition}

\begin{figure}[h]
\begin{centering}
\includegraphics[width=\figwidth]{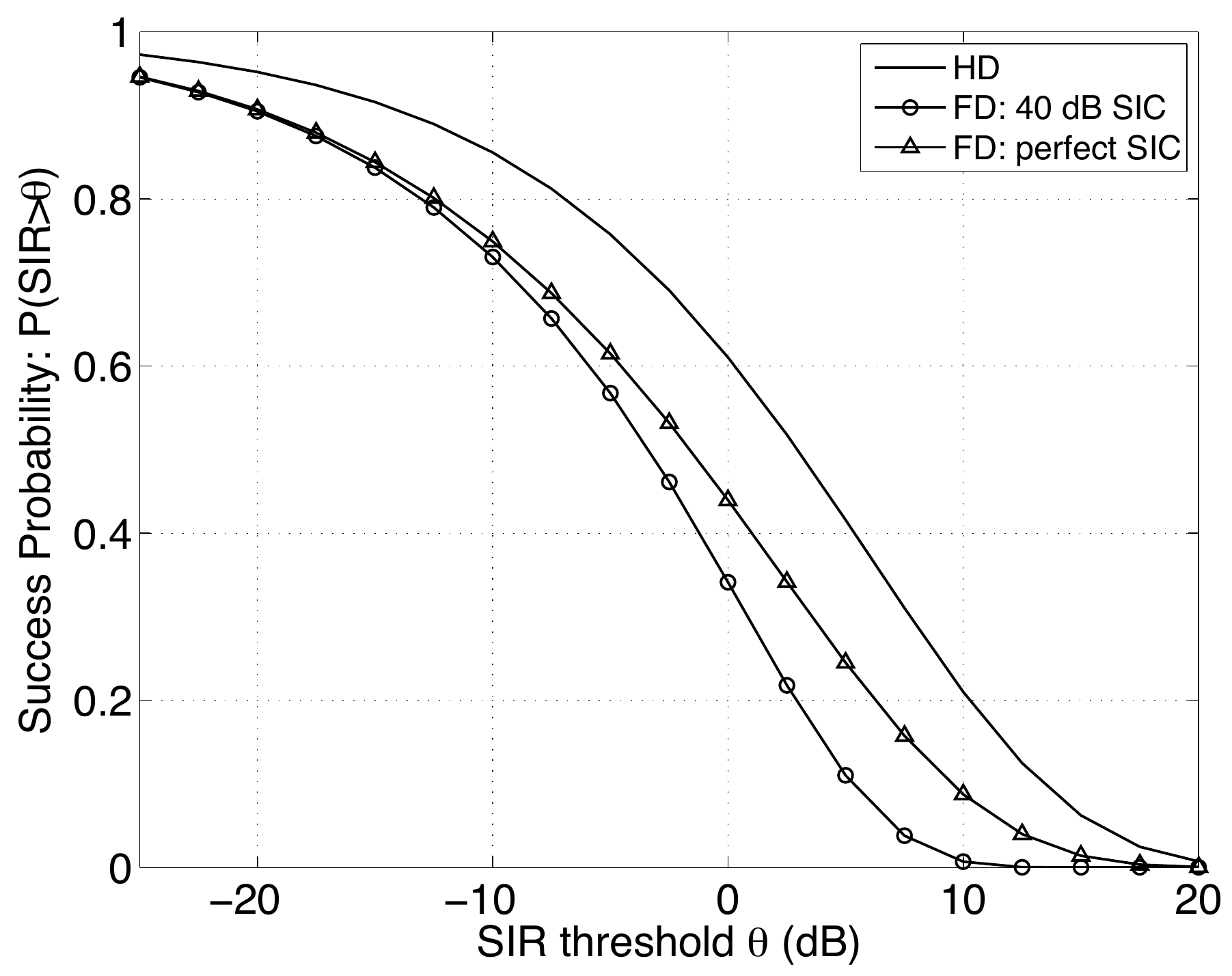}
\par\end{centering}
\caption{Success probabilities of FD-only networks ($p_{s, \;p_2=1}(\theta)$) and HD-only networks ($p_{s, \;p_1=1}(\theta)$) at two SIPRs $\beta = 10^{-4}$ and $\beta=0$. The other parameters are $\alpha=4$, $\lambda=0.1$, $R=1$, $K=-34$ dB (Assume that $G_{\rm tx}=G_{\rm rx}=2$, i.e., $3$ dBi, and $f_c=2.4$ GHz).}
\label{Fig:1-3}
\end{figure}

The following theorem bounds this SIR loss.
\begin{thm} \label{SIR loss}The SIR loss $G(p)$ between the FD-only network and HD-only network is bounded as
\begin{equation}
(1+\delta + \gamma(\theta_{\rm FD}(p)))^{{1}/{\delta}} \leq G(p) \leq (2 + \gamma(\theta_{\rm FD}(p)))^{{1}/{\delta}},\label{SIR_loss}
\end{equation}
where $\gamma(x)= x^{1-\delta}\frac{R^{\alpha-2}\beta \sin(\pi \delta)}{\lambda \pi^2 \delta K}$.
\end{thm}
\begin{IEEEproof}The proof is quite straightforward by equating
\[p_{s, \;p_1=1} (\theta_{\rm HD}) = p_{s, \;p_2=1}(\theta_{\rm FD}),\]
and solving for the ratio $\frac{\theta_{\rm HD}}{\theta_{\rm FD}}$.
From (\ref{ps_HD}) and (\ref{ps_FD}), we obtain
\[\lambda H(\theta_{\rm HD} R^{\alpha},\alpha) =  \lambda F(\theta_{\rm FD} R^{\alpha},\alpha) + \theta_{\rm FD} R^{\alpha} \beta/K,\]
and from (\ref{F2}) in Corollary \ref{cor:F}, we have
\ifCLASSOPTIONonecolumn
\begin{equation}
(1+\delta) \lambda H(\theta_{\rm FD} R^{\alpha},\alpha) \leq \lambda H(\theta_{\rm HD} R^{\alpha},\alpha) - \theta_{\rm FD} R^{\alpha} \beta/K \leq 2\lambda H(\theta_{\rm FD} R^{\alpha},\alpha).\label{F_in}
\end{equation}
\else
\begin{align}
(1+\delta) \lambda H(\theta_{\rm FD} R^{\alpha},\alpha) \leq \lambda H(\theta_{\rm HD} R^{\alpha},\alpha) - \theta_{\rm FD} R^{\alpha} \beta/K \nonumber\\ 
\leq 2\lambda H(\theta_{\rm FD} R^{\alpha},\alpha).\label{F_in}
\end{align}
\fi
By inserting $H(s,\alpha)=\frac{\pi^{2}\delta s^{\delta}}{\sin\left(\pi\delta\right)}$ into (\ref{F_in}) and after elementary manipulations, (\ref{F_in}) leads to 
\begin{equation}
\left(1+\delta + \gamma(\theta_{\rm FD})\right)^{{1}/{\delta}} \leq \frac{\theta_{\rm HD}}{\theta_{\rm FD}} \leq (2 + \gamma(\theta_{\rm FD}))^{{1}/{\delta}}.\label{theta_in}
\end{equation}
Hence, we have (\ref{SIR_loss}).
\end{IEEEproof}
Apparently, the bounds of the SIR loss depend on the SIR threshold $\theta_{\rm FD}$ due to the imperfect self-interference cancellation since $\gamma(\theta_{\rm FD})>0$ for imperfect self-interference cancellation. It means that imperfect self-interference cancellation introduces an extra SIR loss compared to perfect self-interference cancellation and that the SIR loss gets larger as the residual self-interference increases, as shown in Figure \ref{Fig:1-3}.
\begin{cor}\label{cor:G}The SIR loss $G(p)$ between the FD-only network and HD-only network under perfect self-interference cancellation ($\beta =0$) is bounded as
\begin{equation}
(1+\delta)^{{1}/{\delta}} \leq G(p) \leq {2}^{{1}/{\delta}}.\label{SIR_loss2}
\end{equation}
\end{cor}
\begin{IEEEproof}
Follows from Theorem \ref{SIR loss} since $\gamma(\theta_{\rm FD})=0$ for $\beta=0$. 
\end{IEEEproof}
For perfect self-interference cancellation ($\beta=0$), the bounds of the SIR loss only depend on the path loss exponent $\alpha=2/\delta$, i.e., they are independent of the SIR threshold, the target success probability $p$, and the link distance $R$. 
Corollary \ref{cor:G} can also be proven in the following way. Under perfect self-interference cancellation, from (\ref{ps_HD}), we have \[p_{s, \;p_1=1}(\theta_{\rm HD})=e^{-c\theta_{\rm HD}^{\delta}},\]
where $c = \frac{\lambda\pi^{2}\delta R^2}{\sin(\pi\delta)}$.
From (\ref{ps_FD}) and (\ref{F2}), we have 
\begin{align*}
\underline{p}_{s, \;p_2=1}(\theta_{\rm FD})&=e^{-2c\theta_{\rm FD}^{\delta}},\\
\overline{p}_{s, \;p_2=1}(\theta_{\rm FD})&=e^{-(1+\delta)c\theta_{\rm FD}^{\delta}}.
\end{align*}
By solving 
\begin{align*}
\underline{p}_{s, \;p_2=1}(\theta_{\rm FD})&=p_{s, \;p_1=1}(\theta_{\rm FD} \overline{G}),\\
\overline{p}_{s, \;p_2=1}(\theta_{\rm FD})&=p_{s, \;p_1=1}(\theta_{\rm FD} \underline{G}),
\end{align*}
we obtain the upper bound $\overline{G}=2^{\frac{1}{\delta}}$ and the lower bound $\underline{G}=(1+\delta)^{\frac{1}{\delta}}$.

Therefore, the upper (lower) bound of the SIR loss are actually the constant horizontal gap between the upper (lower) bound of the success probability of the FD-only network and that of the HD-only network. Under perfect self-interference cancellation, the upper bound of the success probability of the FD-only network is just the success probability curve of the HD-only network left shifted by  $\underline{G}^{\rm dB}=\frac{10\log_{10}{(1+\delta)}}{\delta}$ dB, whereas the lower bound is that left shifted by $\overline{G}^{\rm dB}=\frac{10\log_{10}{2}}{\delta}$ dB. For $\alpha=4$, the upper bound in Figure \ref{Fig:1-1} is equivalent to the HD curve left-shifted by $3.5$ dB while the lower bound equivalent to that left-shifted by $6.0$ dB.

To summarize, FD-only operation can result in up to $6.0$ dB SIR loss compared to HD-only operation even under perfect self-interference cancellation. Remarkably, this result only depends on the path loss exponent. The above analysis accurately quantifies the SIR loss caused by the extra interference introduced by the FD transmissions. Imperfect self-interference cancellation further adds to the SIR loss, especially when the SIR threshold is high, as shown in \figref{Fig:1-3}.

\section{Throughput Analysis\label{sec:Throughput-Performance-Analysis}}
\subsection{Problem statement}
The purpose of FD transmission in a network is to increase
the network throughput. While FD increases the throughput of an isolated link, it also causes additional interference to the other links. As analyzed in the previous section, FD transmission leads to SIR loss. There is a tradeoff between the link throughput and interference when the nodes in the networks decide to choose FD or HD. Given a network
that consists of nodes of FD capability and HD capability,
how should a node choose between FD and HD operation in order to maximize the network-wide throughput as the network configuration varies? It is important to determine under what condition one should choose FD. 

First, we need to define the throughput. In a
random wireless network described by $\hat{\Phi}$, we can consider the throughput of
the \textit{typical link} as mentioned in the network model.
It has probability $p_{1}$ to be in HD mode and
$p_{2}$ to be in FD mode. Therefore, its throughput can be defined as follows:
\begin{defn}\label{linkT}
For a wireless network described by $\hat{\Phi}$, the throughput is defined as
\begin{equation}
T\triangleq \lambda \left(p_{1}p_{s}^{\rm HD}\log(1+ \theta)+2p_{2}p_{s}^{\rm FD}\log(1+ \theta)\right),\label{eq:T}
\end{equation}
assuming that a spectral efficiency of $\log(1+\theta)$ is achievable for a SIR threshold $\theta$.
\end{defn}

Let $\lambda_{1} \triangleq \lambda p_{1}$ and $\lambda_{2} \triangleq \lambda p_{2}$ be the densities of HD links and FD links. $\lambda_{1}$ and $\lambda_{2}$ can be tuned by changing the transmit probabilities of HD and FD modes $p_1$ and $p_2$ given a fixed node density $\lambda$ or vice versa. By doing so, we can optimize the throughput over the densities of HD and FD links ($\lambda_{1}$ and $\lambda_{2}$) instead of just the transmit probabilities $p_1$ and $p_2$ and reduce the variables by one as well. Hence, (\ref{eq:T}) can be rewritten as 
\begin{equation}
T= \lambda_{1}p_{s}^{\rm HD}\log(1+ \theta)+2\lambda_{2}p_{s}^{\rm FD}\log(1+ \theta).\label{eq:T2}
\end{equation}

Given the definition of throughput, there are two extreme cases that
are particularly relevant: HD-only networks and FD-only networks, as mentioned earlier. Their throughputs are given as

\begin{equation}
T^{\mbox{\scriptsize{\rm HD}}}=\lambda_{1}\log(1+ \theta)\exp(-\lambda_{1}H(\theta_{\rm HD} R^{\alpha},\alpha))\label{HD FD}
\end{equation}
and

\begin{equation}
T^{\mbox{\scriptsize{\rm FD}}}=2\lambda_{2}\kappa\log(1+ \theta)\exp(-\lambda_{2}F(\theta R^{\alpha},\alpha)).\label{HD FD1}
\end{equation}
With the above setup, the goal is to optimize the throughput over the densities $\lambda_1$ and $\lambda_2$:
\begin{equation}
T_{\max} = \max_{\lambda_1,\lambda_2}T(\lambda_{1},\lambda_{2}).
\end{equation}
It is also interesting to find the relationship between the maxima
of $T^{\mbox{\scriptsize{\rm HD}}}$, $T^{\mbox{\scriptsize{\rm FD}}}$ and $T$, denoted as $T_{\max}^{\mbox{\scriptsize{\rm HD}}}$,
$T_{\max}^{\mbox{\scriptsize{\rm FD}}}$ and $T_{\max}$. 

\subsection{Throughput optimization}
Inserting $p_{s}^{\rm HD}$ and $p_{s}^{\rm FD}$ from (\ref{eq:ps-2}) and (\ref{eq:ps-3}) into (\ref{eq:T}), we have
\begin{equation}
T(\lambda_{1},\lambda_{2})= \left(\lambda_{1}+2\lambda_{2}\kappa\right)\exp(-\lambda_{1}H)\exp(-\lambda_{2}F)\log(1+ \theta).\label{eq:T1}
\end{equation}

From now on, we will
use $H$ to denote $H(\theta R^{\alpha},\alpha)$ and $F$
to denote $F(\theta R^{\alpha},\alpha,R)$ for simplicity. $T_{\max}^{\mbox{\scriptsize{\rm HD}}}$ and $T_{\max}^{\mbox{\scriptsize{\rm FD}}}$
can be easily obtained by the following lemma.
\begin{lem}
\label{lem:6}For a HD-only network, described
by $\hat{\Phi}$ with $p_{1}=1$, $T_{\max}^{\mbox{\scriptsize{\rm HD}}}$ is given
by 
\begin{equation}
T_{\max}^{\mbox{\scriptsize{\rm HD}}}=
T^{\mbox{\scriptsize{\rm HD}}}\!\left(\frac{1}{H}\right)=\frac{1}{eH}\log(1+\theta),\label{eq:hdmax}
\end{equation}
with optimal density of HD links
\begin{equation}
\lambda_{1}^{{\scriptsize{\rm opt}}}=\frac{1}{H}.\label{eq:p1opt}
\end{equation}
 For a FD-only network, described by
$\hat{\Phi}$ with $p_{2}=1$, $T_{\max}^{\mbox{\scriptsize{\rm FD}}}$ is given by
\begin{equation}
T_{\max}^{\mbox{\scriptsize{\rm FD}}}=
T^{\mbox{\scriptsize{\rm FD}}}\!\left(\frac{1}{F}\right)=\frac{2}{e\kappa F}\log(1+\theta),\label{eq:fdmax}
\end{equation}
with optimal density of FD links
\begin{equation}
\lambda_{2}^{{\scriptsize{\rm opt}}}=\frac{1}{F}.\label{eq:p2opt}
\end{equation}
\end{lem}
\begin{IEEEproof}
The proof is straightforward by taking the derivatives of $T^{\mbox{\scriptsize{\rm HD}}}$
and $T^{\mbox{\scriptsize{\rm FD}}}$ with respect to $\lambda_{1}$ and $\lambda_{2}$, respectively.
\end{IEEEproof}
A similar result for HD-only networks has been presented in \cite[Proposition 4]{Haenggi09twc}. In fact, ${1}/{H}$ and ${1}/{F}$ are the spatial efficiency~\cite{Haenggi09twc} of HD-only networks and FD-only networks, respectively. The spatial efficiency quantifies how efficiently a wireless network uses space as a resource. A large spatial efficiency indicates high spatial reuse.

In the following theorem, we show that $T_{\max}$ is achieved by
setting all concurrently transmitting nodes to be in FD mode or in HD mode or in a mixed FD/HD mode, depending on one simple condition. 
\begin{thm} Let 
\begin{equation} L \triangleq\{(\lambda_1, \lambda_2) \in (\mathbb{R}^+)^2\colon \lambda_1 + 2\kappa\lambda_2 = H^{-1}\}.\label{condition}\end{equation}
For a wireless network described by $\hat{\Phi}$, the maximal throughput
is given by 
\begin{equation}
T_{\max} = \begin{cases}
T_{\max}^{\mbox{\scriptsize{\rm FD}}} & \mbox{if } F < 2\kappa H\\
T_{\max}^{\mbox{\scriptsize{\rm HD}}}  & \mbox{if } F > 2\kappa H\\
T_{\max}^{\mbox{\scriptsize{\rm HD}}}= T_{\max}^{\mbox{\scriptsize{\rm FD}}}  & \mbox{if } F = 2\kappa H
\end{cases}\label{eq:equ}
\end{equation}
with the optimal densities of HD and FD links 
\begin{equation}
\left(\lambda_{1}^{{\scriptsize{\rm opt}}},\lambda_{2}^{{\scriptsize{\rm opt}}}\right)\begin{cases}
=\left(0, \frac{1}{F}\right) & \mbox{if } F < 2\kappa H\\
=\left(\frac{1}{H}, 0\right) & \mbox{if } F > 2\kappa H\\
\in L & \mbox{if } F = 2\kappa H. \end{cases}
\end{equation}
\end{thm}
\begin{IEEEproof}
Taking the derivative of $T$ w.r.t. $\lambda_{1}$ and $\lambda_{2}$ leads to 
\begin{equation}
\frac{\partial T}{\partial \lambda_{1}}=\exp(-\lambda_{1}H-\lambda_{2}F)\log(1+\theta)[1- H(2\kappa \lambda_{2}+\lambda_{1})],\label{eq:p1}
\end{equation}
\begin{equation}
\frac{\partial T}{\partial \lambda_{2}}=\exp(-\lambda_{1}H-\lambda_{2}F)\log(1+\theta)[2\kappa-F(2\kappa\lambda_{2}+\lambda_{1})].\label{eq:p2}
\end{equation}
Setting $\frac{\partial T}{\partial \lambda_{1}}=0$ and $\frac{\partial T}{\partial \lambda_{2}}=0$, we have
\begin{equation}
\lambda_{1}+2\kappa\lambda_{2} = \frac{1}{H}\label{condition1}
\end{equation}
\begin{equation}
\lambda_{1}+2\kappa\lambda_{2} = \frac{2\kappa}{F}.\label{condition2}
\end{equation}

\leavevmode
\begin{enumerate} \item $F= 2\kappa H$: Both partial derivatives are zero as long as  $(\lambda_1, \lambda_2)\in L$. $({1}/{H}, 0)$ and $(0, {2\kappa}/{F})$ both lie in $L$. Hence, $T_{\max} =  T({1}/{H}, 0) = T_{\max}^{\mbox{\scriptsize{HD}}}$ and also $T_{\max} = T(0, {2\kappa}/{F}) = T_{\max}^{\mbox{\scriptsize{FD}}}$. This case is the break-even point where FD and HD have the same throughput. That means in a wireless network, the typical link has the same throughput no matter if it is in FD or HD mode.

\item $F > 2\kappa H$: Under this condition, let $\frac{\partial T}{\partial \lambda_{1}}=0$ and we have (\ref{condition1}). Moreover, $\frac{\partial T}{\partial \lambda_{2}}<0.$ Therefore, the maximal $T$ is achieved at $\left(\lambda_{1},\lambda_{2}\right)=\left({1}/{H}, 0\right)$ from (\ref{condition1}) and $\frac{\partial T}{\partial \lambda_{2}}<0$. Note that $T({1}/{H}, 0)=T^{\mbox{\scriptsize{HD}}}({1}/{H})=T^{\mbox{\scriptsize{HD}}}_{\max}$. On the other hand, letting  $\frac{\partial T}{\partial \lambda_{2}}=0$, we have (\ref{condition2}) and $\frac{\partial T}{\partial \lambda_{1}}>0,$ which leads to that the maximal $T$ is achieved at $\left(\lambda_{1},\lambda_{2}\right)=({2\kappa}/{F}, 0).$ Since $T({2\kappa}/{F}, 0)<T({1}/{H}, 0)$. We conclude that $T_{\max} = T^{\mbox{\scriptsize{HD}}}$ in this case.
\item $F < 2\kappa H$: By similar reasoning as in the second case, we can conclude that $T_{\max} = T^{\mbox{\scriptsize{FD}}}$ in this case.
\end{enumerate}\end{IEEEproof}
Under perfect self-interference cancellation ($\beta = 0$), $F < 2\kappa G$, and we always have 
\[
T_{\max}=T_{\max}^{\mbox{\scriptsize{FD}}},
\] which means $T_{\max}$ is always achieved by
setting all transmitting nodes to work in FD mode, despite the extra interference
caused by the FD nodes. This conclusion holds for all network
configurations $(\lambda, \theta, R, \alpha)$.

On the other hand, the following corollary quantifies how the imperfect self-interference cancellation affects the throughput in a wireless network with full-duplex radios.
\begin{cor}\label{beta_c}
Given an SIR threshold $\theta$, path loss exponent $\alpha$, and link distance $R$, there exists a critical SIPR value $\beta_{c}$ in the wireless network described by $\hat{\Phi}$: when $\beta < \beta_{c}$, FD is preferable in terms of throughput while HD has better throughput when $\beta > \beta_{c}$, where 
\begin{equation}
\beta_{c} = \frac{K\log({2H}/{F})}{\theta R^{\alpha}}.\label{beta}
\end{equation}
\end{cor}
\begin{IEEEproof}
$\beta_{c}$ can be obtained by solving $F=2\kappa H = 2e^{-\theta R^{\alpha} \beta/K}H$.
By Corollary \ref{cor:F}, the ratio of $F/H$ does not depend on $R$ and hence $\beta_{c}$ scales as $R^{-\alpha}$.
\end{IEEEproof}

\begin{figure}[h]
\begin{centering}
\includegraphics[width=\figwidth]{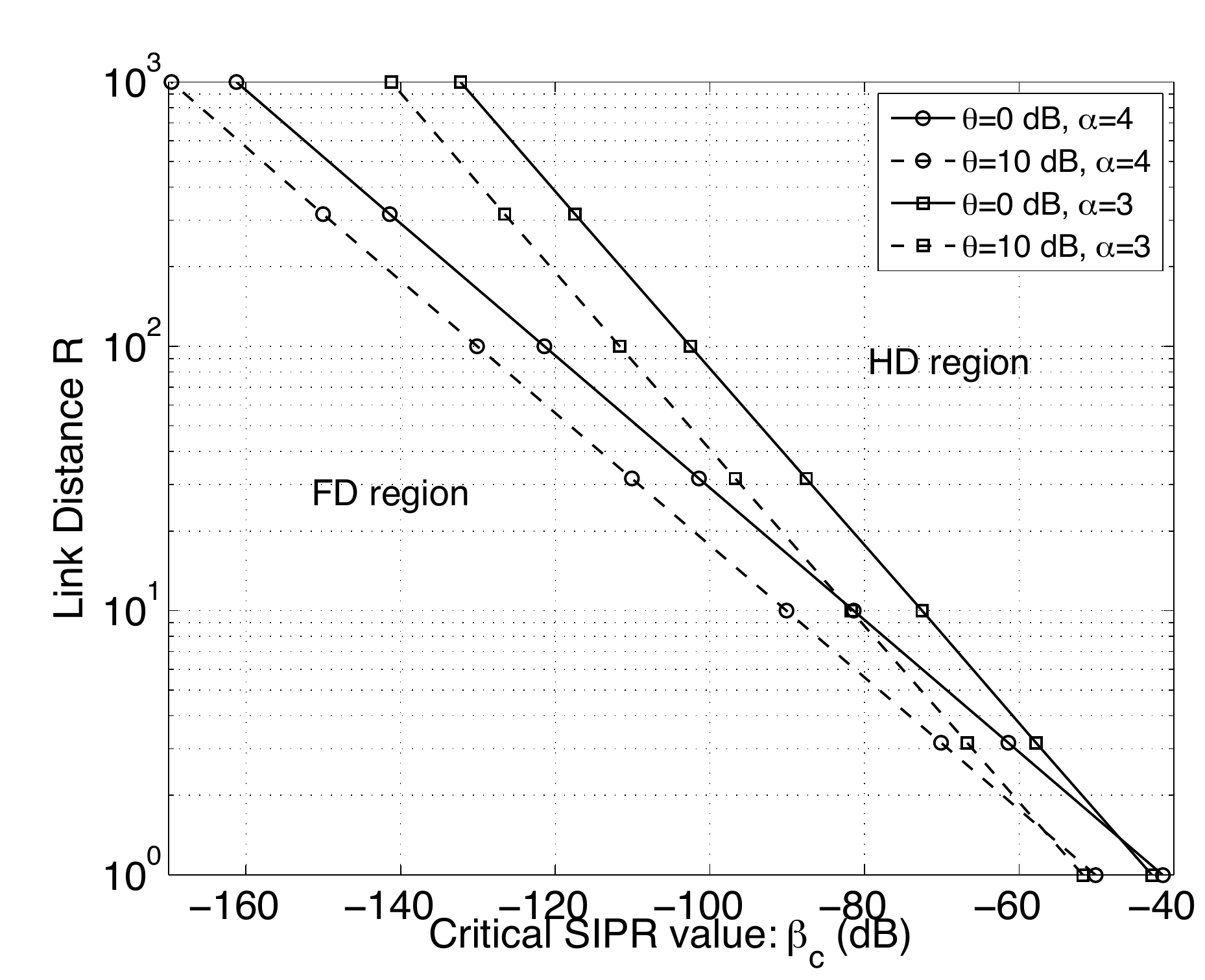}
\par\end{centering}
\caption{The link distance vs the critical SIPR $\beta_{c}$ from (\ref{beta}) with $K=-34$ dB for $G_{\rm tx}=G_{\rm rx}=2$, i.e., $3$ dBi, and $f_c=2.4$ GHz, which corresponds to the carrier frequency of a WiFi signal. Below the curves, FD provides a higher throughput, while above the curves, HD does.}
\label{Fig:2}
\end{figure}
Figure \ref{Fig:2} plots the relationship between the link distance and the self-interference cancellation threshold $\beta_{c}$. It provides very valuable insight into the system design. The curves are linear in this log-log plot with slope $-1/\alpha$. The region under the lines is the region where FD transmission achieves a higher network throughput. For example, assume that the self-interference cancellation is limited to $80$ dB due to hardware imperfection. In this case, FD transmission is preferable only if the link distance is smaller than $10$ when $\theta=0$ dB and $\alpha = 4$ under the wireless network model used in this paper. To achieve a link distance of up to $100$, the self-interference cancellation needs to be at least $100$ dB ($\alpha = 3$) and $120$ dB ($\alpha = 4$) when $\theta =0$ dB. When the link distance is greater than $100$ with self-interference cancellation no greater than $120$ dB, it is better to use HD. So the amount of self-interference cancellation determines the maximal transmission range for which FD has better throughput than HD.

\subsection{Comparison of FD with HD}
Since the mixed FD/HD network achieves the maximal throughput in
the extreme case of a FD-only or HD-only network, we can simply focus on FD-only
and HD-only networks and compare their maximal throughputs
from the results in Lemma \ref{lem:6}. 
The throughput gain of a FD network over a HD network is of
great interest. It is defined as follows.
\begin{defn}
The throughput gain (TG) is defined as the ratio between the maximal
throughput of FD-only networks and HD-only networks given the same network parameters
$\left(\theta,R,\alpha\right)$:
\[
{\rm TG}\triangleq\frac{T_{\max}^{\mbox{\scriptsize{FD}}}}{T_{\max}^{\mbox{\scriptsize{HD}}}}.
\]
\end{defn}
The following corollary quantifies and bounds $\rm{TG}$ in
terms of $F$ and $H$. Note that $F$ and $H$ are constant given
$\left(\theta,R,\alpha\right)$.
\begin{cor}
The throughput gain is given by 
\begin{equation}
{\rm TG}=\frac{2\kappa H}{F}\label{eq:TG}
\end{equation} and bounded as
\begin{equation}
\kappa  < {\rm TG} < \frac{2\kappa}{1+\delta}.
\label{eq:TG_ub}
\end{equation}
Moreover, for any $\beta \geq 0$, 
\begin{equation}{\rm TG}(\theta) \sim \frac{2\kappa}{1+\delta}=\frac{2}{1+\delta}\exp(-\theta R^\alpha\beta/K), \quad\theta\to\infty.\label{TGinf}\end{equation}
\end{cor}
\begin{IEEEproof}
From (\ref{eq:hdmax}) and (\ref{eq:fdmax}), we have (\ref{eq:TG}). The upper and lower bounds are easily obtained from Corollary \ref{cor:F}. 

For $\beta>0$, 
\begin{equation}\lim_{\theta \rightarrow \infty} \kappa = \lim_{\theta \rightarrow \infty} e^{-\frac{\theta R^{\alpha} \beta}{K}}=0.\end{equation}
Therefore, both the lower bound and upper bound of  ${\rm TG}$ converge to $0$ as $\theta$ goes to $\infty$. As a result, \begin{equation}\lim_{\theta \rightarrow \infty} {\rm TG}(\theta)=0.\label{TGinf1}\end{equation} $\beta=0$ implies $\kappa = 1$, which leads to \begin{equation}\lim_{\theta \rightarrow \infty} {\rm TG}(\theta)=\lim_{\theta \rightarrow \infty} \frac{2H}{F}=\frac{2}{1+\delta}\label{TGinf2}\end{equation}
 from (\ref{FH2}). Combining (\ref{TGinf1}) and (\ref{TGinf2}), we obtain (\ref{TGinf}).
\end{IEEEproof}

(\ref{TGinf}) indicates that the throughput gain converges to its upper bound as $\theta$ goes to infinity. 
\ifCLASSOPTIONonecolumn
\begin{figure}[h]
\begin{centering}
\subfigure[$50$ dB self-interference cancellation: $\beta=10^{-5}$]{\includegraphics[width=7cm]{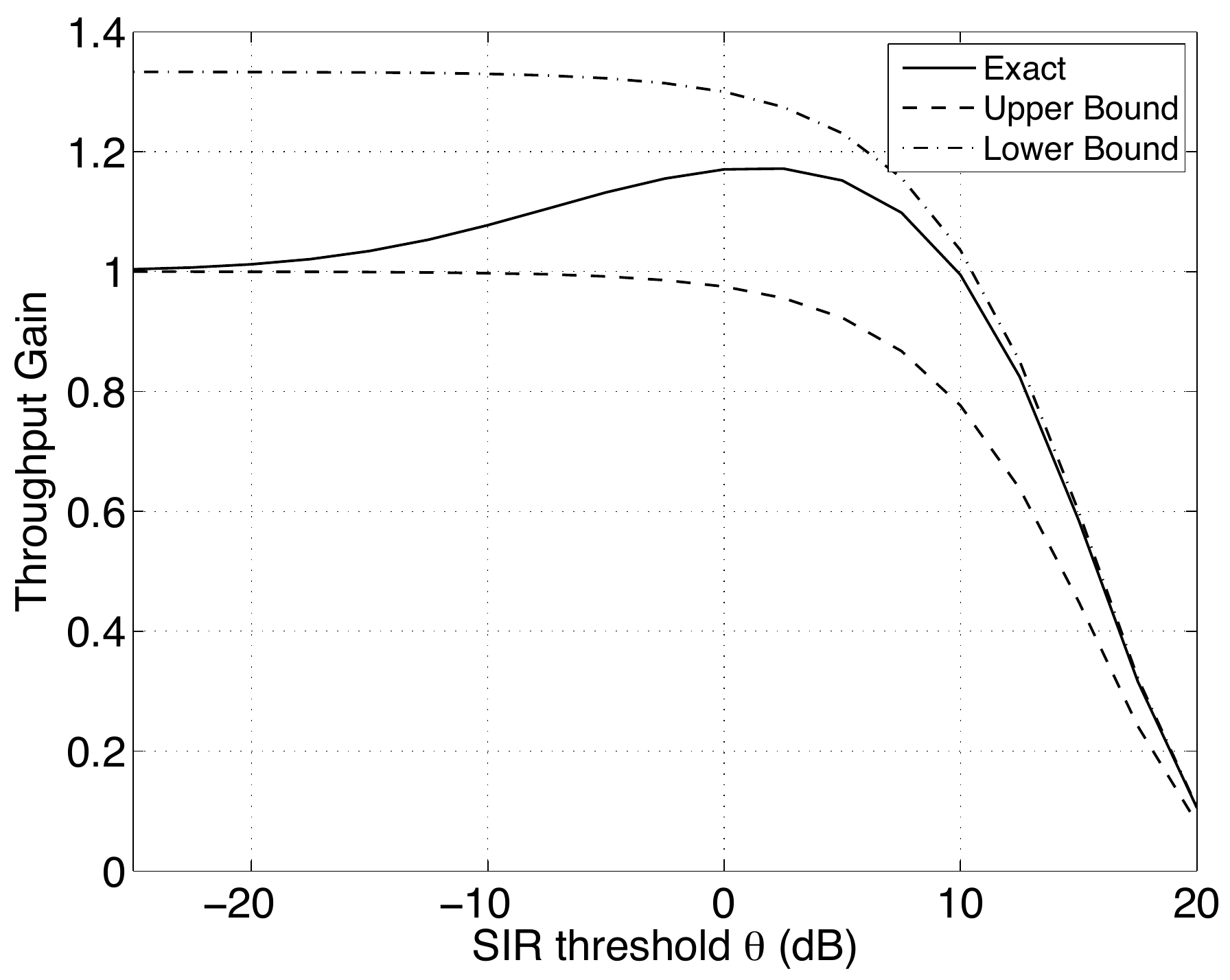}\label{beta40}
}
\subfigure[$70$ dB self-interference cancellation: $\beta=10^{-7}$]{\includegraphics[width=7cm]{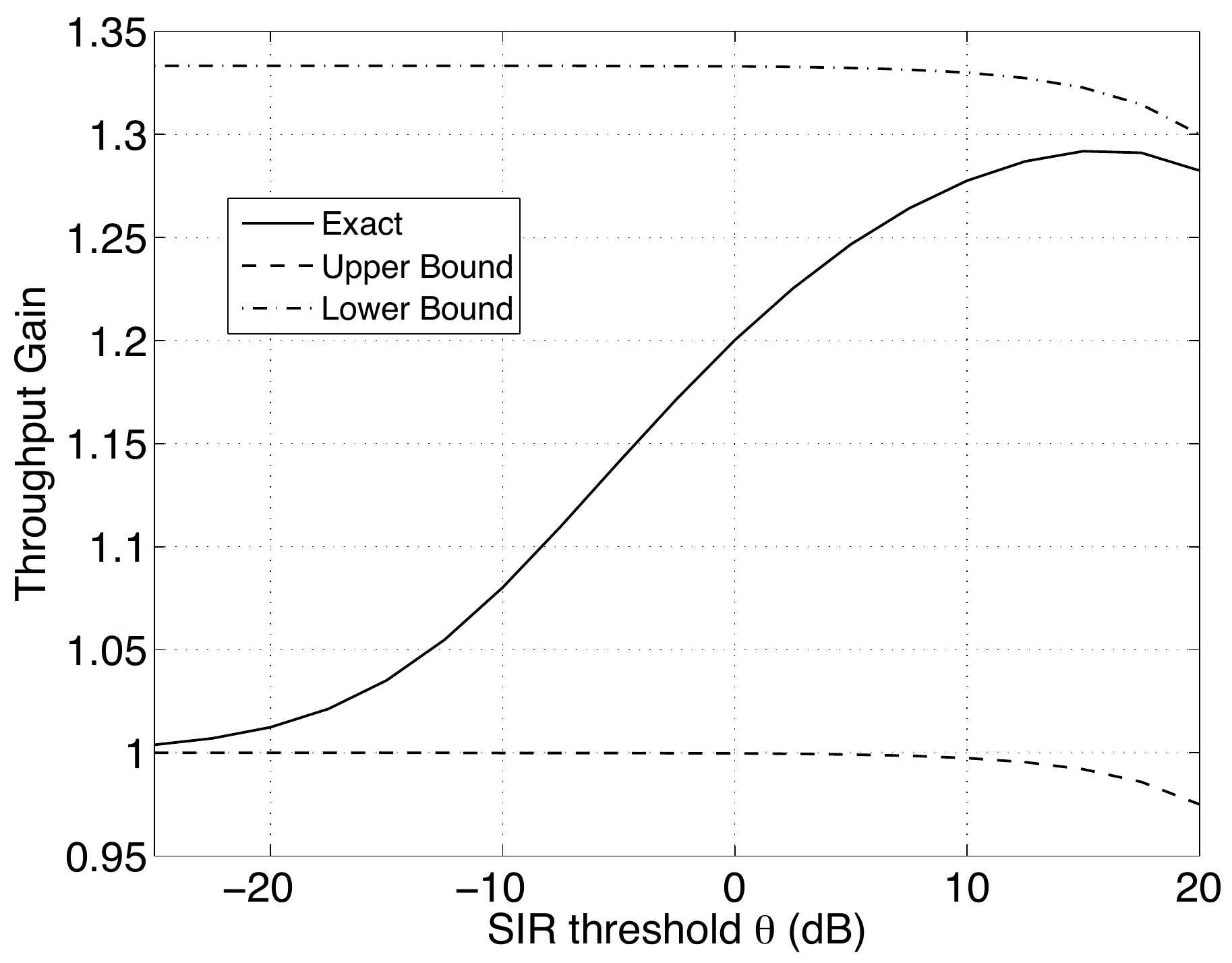}\label{beta20}
}
\par\end{centering}
\begin{centering}
\subfigure[Perfect self-interference cancellation: $\beta=0$]{\includegraphics[width=7cm]{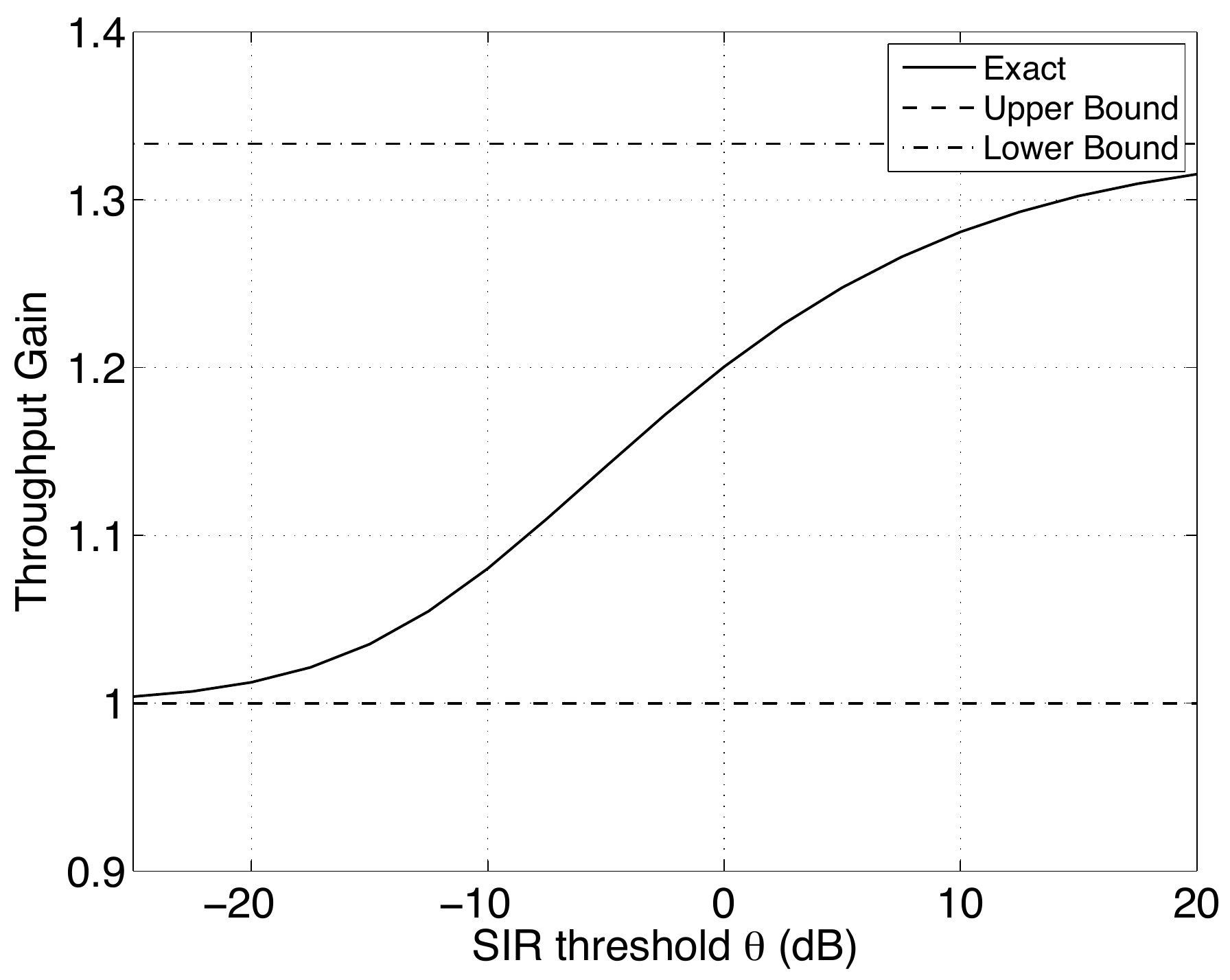}\label{beta0}
}
\par\end{centering}
\caption{Throughput gain as a function of the SIR threshold $\theta$ and its bounds at different SIPRs for
$\alpha=4$, $R=1$.}
\label{Fig:4}
\end{figure}
\else
\begin{figure}[h]
\begin{centering}
\subfigure[$50$ dB self-interference cancellation: $\beta=10^{-5}$]{\includegraphics[width=7cm]{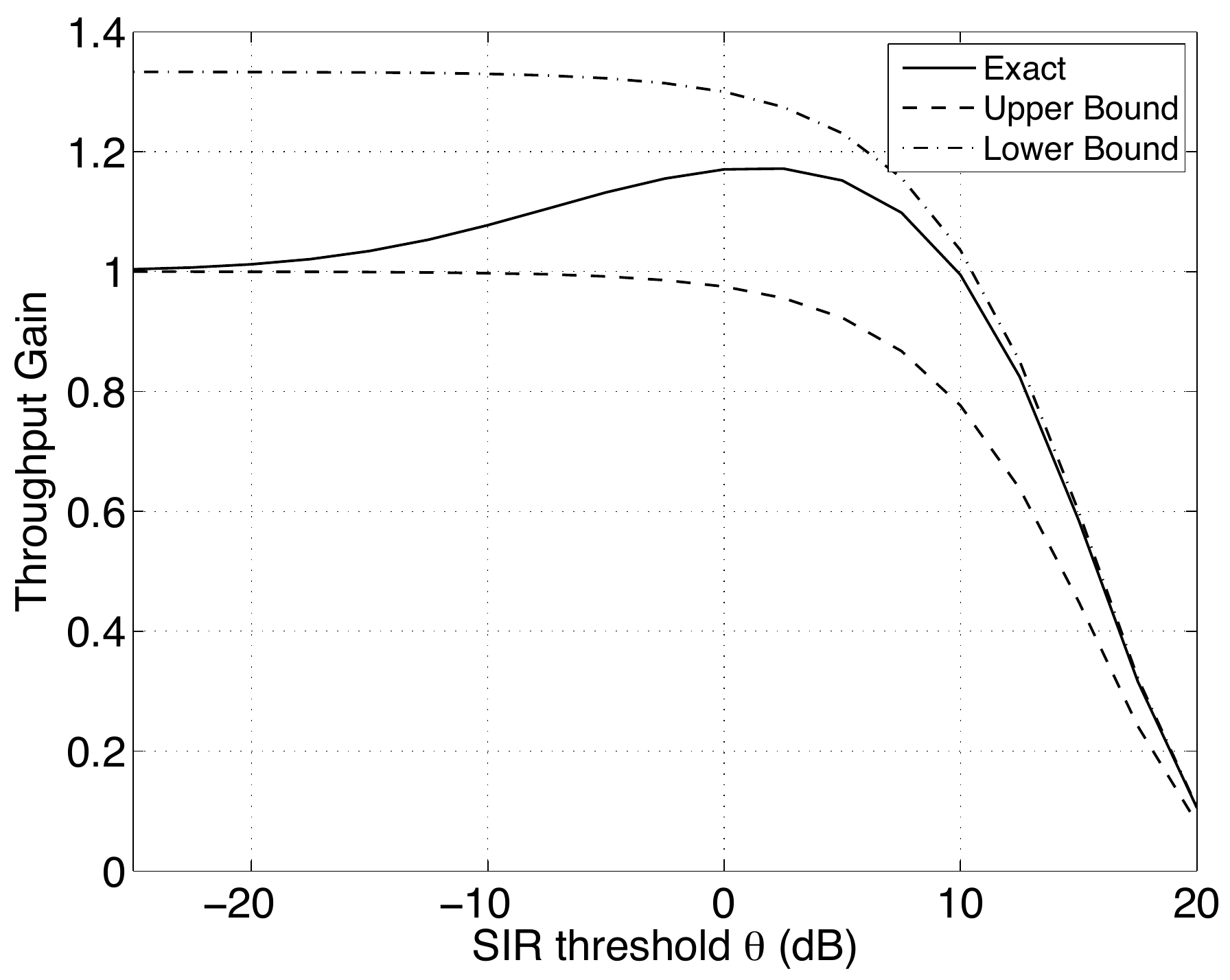}\label{beta40}
}
\par\end{centering}
\begin{centering}
\subfigure[$70$ dB self-interference cancellation: $\beta=10^{-7}$]{\includegraphics[width=7cm]{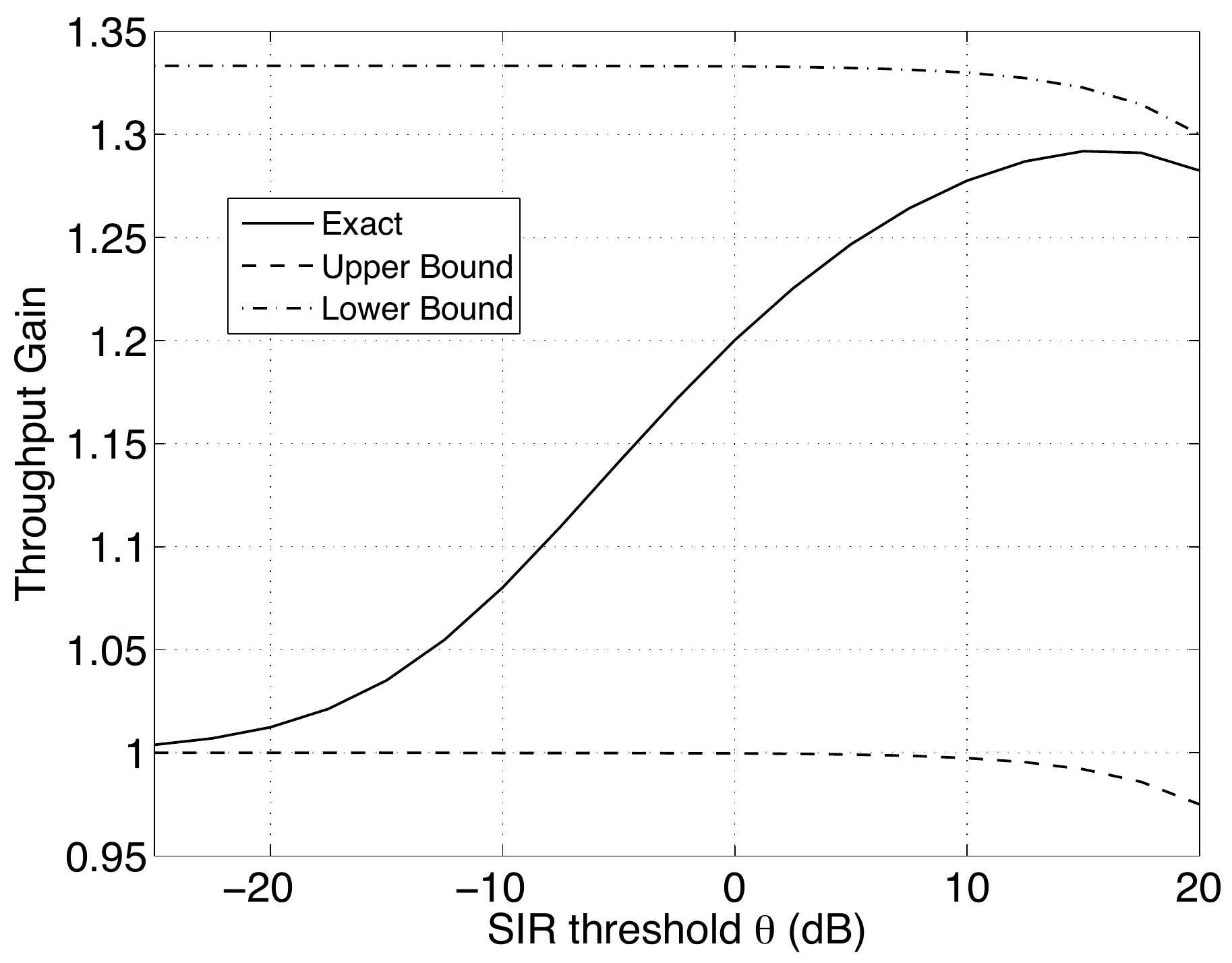}\label{beta20}
}
\par\end{centering}
\begin{centering}
\subfigure[Perfect self-interference cancellation: $\beta=0$]{\includegraphics[width=7cm]{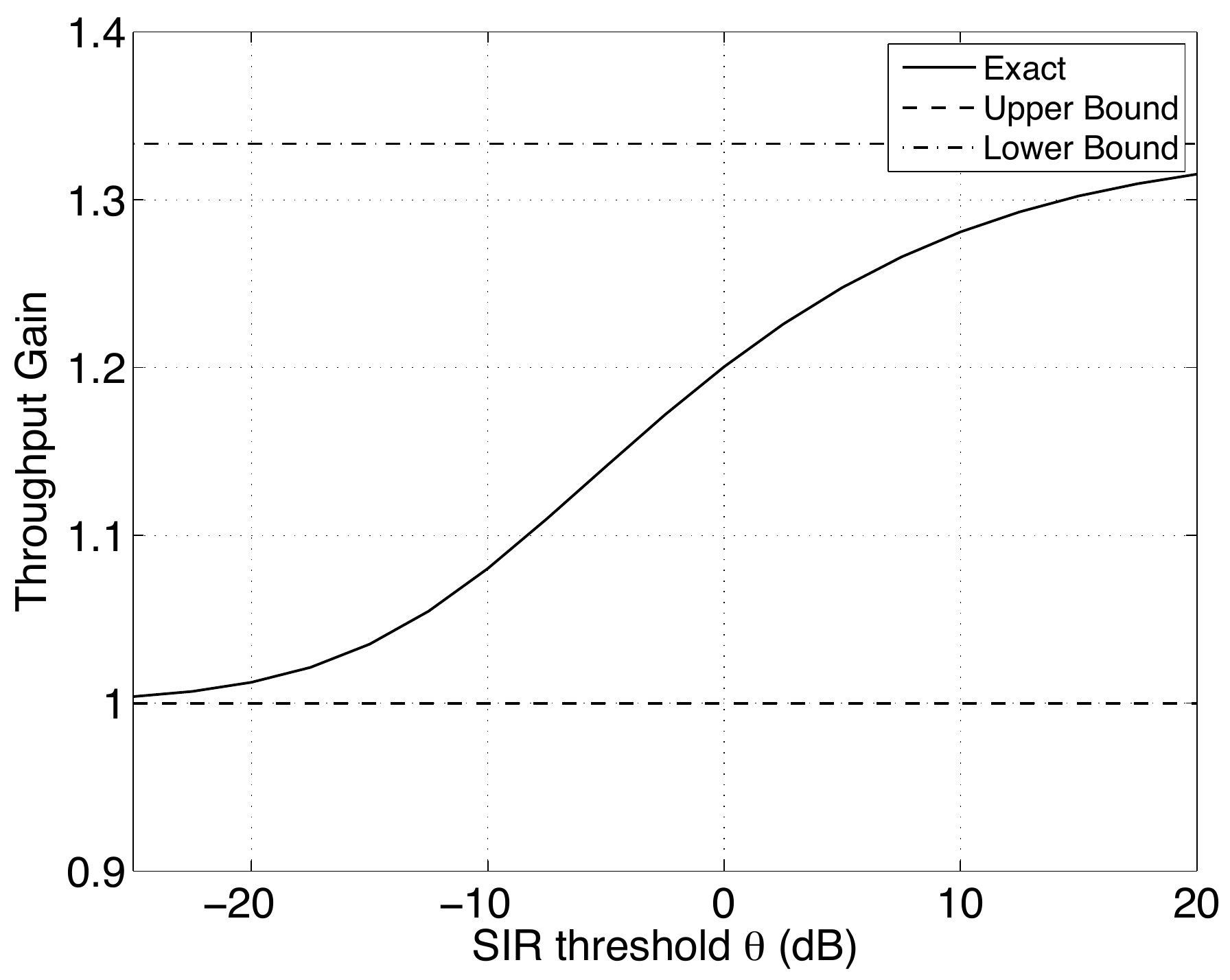}\label{beta0}
}
\par\end{centering}
\caption{Throughput gain as a function of the SIR threshold $\theta$ and its bounds at different SIPRs for
$\alpha=4$, $R=1$.}
\label{Fig:4}
\end{figure}
\fi
\figref{Fig:4} illustrates the throughput gain as a function
of the SIR threshold together with its upper and lower bounds given in (\ref{eq:TG_ub}). As seen, the throughput gain
is always lower than $\frac{2}{1+\delta}$ since the upper bound is smaller than that. For perfect self-interference cancellation, the throughput gain increases as the SIR threshold gets larger as shown in \figref{beta0}. The throughput gain decreases in the high SIR regime due to the imperfect self-interference cancellation as shown in \figref{beta40} and \figref{beta20}. When the self-interference cancellation is not sufficient, the throughput gain is less than $1$, which means that the HD-only network has a higher throughput, i.e., the value in the curve after $\theta>10$ dB in \figref{beta40} is less than $1$. These figures illustrate the throughput gain under different conditions, especially the impact of the imperfect self-interference cancellation on the throughput.

\section{Conclusion\label{sec:Conclusion}}

In this paper, we analyzed the throughput of wireless networks with
FD radios using tools from stochastic geometry. Given a wireless network of radios
with both FD and HD capabilities, we showed that FD transmission is always preferable
compared to HD transmission in terms of throughput when the self-interference cancellation is perfect. It turns out that the throughput of HD transmission cannot be doubled and the actual gain is $\frac{2\alpha}{\alpha+2}$ for an ALOHA protocol, where $\alpha$ is the path loss exponent. Under imperfect self-interference cancellation, the network has a break-even point where FD and HD have the same throughput. The break-even point depends on the amount of self-interference cancellation and the link distance. Given a fixed SIR threshold and path loss exponent, the necessary amount of self-interference cancellation in dB is logarithmically proportional to the link distance. It means that the residual self-interference determines the maximal link distance within which FD is beneficial compared to HD. It provides great insights into the network design with FD radios. We also analyzed and quantified the effects of imperfect self-interference cancellation on the network throughput and SIR loss. The SIR loss of a FD-only network over a HD-only network is quantified within tight bounds. The horizontal gap is utilized to determine the SIR loss. Moreover, the throughput gain of FD over HD is presented under imperfect self-interference cancellation. 

In our network model, we consider the interference-limited case where the thermal noise is ignored. However, this is not a restriction as the throughput gain is independent of the thermal noise since the thermal noise adds the same exponential factor ($\exp({-\theta R^{\alpha} W})$ \cite[Page 105]{Haenggi12book}, where $W$ is the thermal noise power) to the throughput expressions of both HD and FD networks and they cancel each other in the throughput gain.

In general, FD is a
very powerful technique that can be adapted for the next-generation
wireless networks. The throughput gain may be larger if more advanced
MAC protocols other than ALOHA are used or the interference management
can be used for the pairwise interferers in the FD links. A FD-friendly MAC scheme should let the node decide to use FD or HD based on its surrounding interference in order to maximize the overall network throughput. There is a strong need for a MAC protocol tailored for a wireless network of radios with both FD and HD capacities and an intelligent and adaptive scheme to switch between FD and HD based on different network configurations. 

\section*{Acknowledgment} 
 This work has been partially supported by the U.S. NSF (grants ECCS-1231806, CNS 1016742 and CCF 1216407).

\bibliographystyle{IEEEtran}
\bibliography{reference}

\end{document}